\documentclass[
  aps,
  prd,
  reprint,
  superscriptaddress,
  nofootinbib,
  floatfix
]{revtex4-2}
\usepackage{amsmath}
\allowdisplaybreaks[4]
\usepackage[T1]{fontenc}
\usepackage{amsmath,amssymb}
\usepackage{graphicx}
\usepackage{dcolumn}
\usepackage{bm}
\usepackage{makecell}
\usepackage{xcolor}
\usepackage{hyperref}
\usepackage{extarrows}
\hypersetup{
  colorlinks=true,
  linkcolor=blue,
  citecolor=blue,
  urlcolor=blue
}

\colorlet{RED}{red}

\begin{document}

\title{\boldmath Extreme mass-ratio inspirals around rotating accelerating black holes}

\author{Xin-Dong Du}
\affiliation{School of Physics and Optoelectronics, South China University of Technology, Guangzhou 510641, China}

\author{Tao Zhou}
\affiliation{School of Physics and Optoelectronics, South China University of Technology, Guangzhou 510641, China}

\author{Wei Xiong}
\affiliation{School of Physics and Optoelectronics, South China University of Technology, Guangzhou 510641, China}

\author{Tieguang Zi}
\email{zitieguang@ncu.edu.cn}
\affiliation{Department of Physics, Nanchang University, Nanchang 330031, China}
\affiliation{Center for Relativistic Astrophysics and High Energy Physics, Nanchang University, Nanchang 330031, China}

\author{Peng-Cheng Li}
\email{pchli2021@scut.edu.cn}
\affiliation{School of Physics and Optoelectronics, South China University of Technology, Guangzhou 510641, China}

\date{\today}

\begin{abstract}
Extreme mass-ratio inspirals (EMRIs) can magnify small departures from Kerr dynamics into appreciable gravitational-wave phase shifts accumulated over many orbital cycles. We exploit this sensitivity to investigate the imprint of a rotating black hole's acceleration on an EMRI waveform. The spinning C metric poses two obstacles to the standard Kerr flux framework: the spacetime is not asymptotically flat, and the acceleration breaks the reflection symmetry that supports exactly equatorial circular timelike orbits. For sufficiently small acceleration $AM$, we therefore formulate the calculation in an intermediate Kerr-like wave zone satisfying $M/r\ll1$ and $Ar\ll1$, and construct a near-equatorial circular orbit by examining its coupled radial--polar stability. We derive the separated point-particle source for the spin$-2$ radial Teukolsky equation, construct a regular normalized angular solution, solve the radial equation using the Sasaki--Nakamura transformation and the Green function method, and couple the resulting horizon and far-zone fluxes to the adiabatic evolution of stable near-equatorial circular orbits. The framework recovers the Kerr limit and reproduces the dominant $l=2$ Kerr fluxes with relative errors of order $10^{-7}$. Acceleration modifies both radiation reaction and the orbital frequency, producing a characteristic nonmonotonic accumulated dephasing. For $M=10^6M_\odot$, $m_s/M=10^{-5}$, $a/M=0.7$, and $AM=3\times10^{-7}$, the dominant-mode dephasing slightly exceeds $1$ rad over one year. Thus even weak acceleration can generate an order-radian secular phase imprint on long-duration EMRIs within the controlled regime of the present approximation.
\end{abstract}

\maketitle

\section{Introduction}\label{sec:1}
Extreme mass-ratio inspirals (EMRIs) are compact binaries in which a stellar-mass object gradually inspirals into a massive black hole, typically with a mass ratio of $10^{-4}$--$10^{-7}$ \cite{LIGOScientific:2016aoc,Amaro-Seoane:2012lgq,Babak:2017tow}. During the final years before plunge, the secondary can complete about $10^5$ relativistic orbital cycles and produce a long-lived, information-rich signal in the millihertz band. EMRIs are therefore primary targets of planned space-based gravitational-wave observatories, including LISA~\cite{LISA:2017pwj}, TianQin~\cite{TianQin:2020hid}, and Taiji \cite{Ruan:2018tsw}. Because the secondary repeatedly probes the strong-field region of the primary, the accumulated waveform phase encodes the central object's multipolar structure with exceptional precision. EMRI observations can consequently map black-hole spacetimes and test general relativity, the Kerr hypothesis, and the black-hole no-hair relation in the strong-field regime~\cite{Ryan:1995wh,Glampedakis:2005cf,Barack:2006pq,Moore:2017lxy,Cardoso:2019rvt,Zi:2021pdp,Zi:2023qfk,Zi:2023geb,Datta:2024vll,Fu:2024cfk,Cardenas-Avendano:2024mqp,Zi:2024jla,Babichev:2024hjf,Kumar:2025jsi,Zare:2025aek,LaHaye:2025ley,Zhao:2025sck,Lu:2025xlp,Xia:2026aty,Long:2026dcb,Muguruza:2026hqn}. Beyond probing the central spacetime, these observations can search for additional fields and environmental effects~\cite{Kocsis:2011dr,Barausse:2014tra,Maselli:2020zgv,LISA:2022kgy,Barsanti:2022ana,Cardoso:2022whc,Speri:2022upm,Zi:2022hcc,Dai:2023cft,Zi:2023omh,Zhao:2024bpp,Jiang:2024lwg,Mitra:2025tag,Vicente:2025gsg,Wang:2025rrj,Dyson:2025dlj,Polcar:2025yto,Kejriwal:2025jao,Luo:2025ewp,Copparoni:2025vty,Haroon:2025rzx,Das:2025vja,Das:2025eiv,Rahman:2025mip,HegadeKR:2025dur,Zhao:2026yis,Zi:2026zpw,Azreg-Ainou:2026xcc,Hu:2026way,Fu:2026gcu}, help distinguish among different EMRI formation channels~\cite{Pan:2021oob,Peng:2022hqa,Pan:2021ksp,LISA:2022yao,Naoz:2022rru}, and serve as cosmological probes that independently constrain the cosmic expansion history~\cite{MacLeod:2007jd,Laghi:2021pqk,Liu:2023onj,LISACosmologyWorkingGroup:2022jok,toscani2024strongly,Zhu:2024qpp}. This broad scientific potential, together with the cumulative phase sensitivity of long-duration EMRI signals, motivates extending accurate waveform modeling beyond the standard assumption of an isolated Kerr primary.

Most EMRI studies assume that MBH is an isolated Kerr black hole whose center of mass follows an inertial trajectory. It is nevertheless useful to ask how the signal changes when the primary itself accelerates~\cite{CalderonBustillo:2018zuq,Woodford:2019tlo}. Anisotropic gravitational-wave emission in a black-hole merger can impart a short-duration acceleration, including the ``superkick'' configurations found in numerical relativity \cite{Merritt:2004xa,Bruegmann:2007bri}, while cosmic strings or external fields can provide a more persistent acceleration and can generate accelerating black-hole geometries \cite{Bousso:1996au,Hawking:1995zn}. The spinning C metric \cite{Plebanski:1976gy,Griffiths:2005se} describes a class of rotating accelerating black holes, and it supplies a simple exact model in which the gravitational consequences of black-hole acceleration can be isolated. This is conceptually distinct from studies of the kinematic motion of an entire EMRI source relative to the detector, where aberration and Doppler effects excite additional waveform modes without modifying the local strong-field geometry \cite{Torres-Orjuela:2020oxq,Torres-Orjuela:2020dhw}: here the acceleration is encoded directly in the background spacetime and alters the orbit, the perturbation equations, and the radiation reaction. The spinning C metric is also theoretically attractive as an exact vacuum, which is a Petrov type-D generalization of Kerr. Its massless-field perturbation equation separates completely into radial and angular sectors \cite{Bini:2008mzd}, making it a rare tractable setting for applying black-hole perturbation theory beyond Kerr. Previous studies have examined the quasinormal modes of accelerating black holes, including a distinct family of acceleration modes \cite{Xiong:2023usm,Chen:2024rov,ZhouLi2025,Han:2024rus}, as well as their photon regions, shadows, lensing signatures, and differential time delays \cite{Grenzebach2015,ZhangJiang2021,Ashoorioon2022}. The gravitational radiation and adiabatic evolution of EMRIs in this background, however, have not yet been developed.

In this work, we construct a small-acceleration framework for computing EMRI radiation from a rotating accelerating black hole. Starting from the inhomogeneous spin-$-2$ Teukolsky equation, we derive the separated point-particle source, determine a regular normalized angular solution, and solve the radial equation by combining the Sasaki--Nakamura transformation with a Green-function construction. Because acceleration breaks north--south reflection symmetry, we construct a near-equatorial circular orbit, and analyze its coupled radial--polar stability instead of assuming Kerr-like equatorial motion. A second complication is that the acceleration horizon removes the usual asymptotically flat wave zone. For $AM\ll1$, we introduce an intermediate Kerr-like region $r_+\ll r_\infty\ll r_A$, impose an approximate outgoing-wave condition there, and evaluate the far-zone flux. We then use flux balance to evolve the orbit and accumulate the gravitational-wave phase. The calculation recovers the Kerr limit, reproduces the dominant $l=2$ Kerr fluxes with relative errors of order $10^{-7}$, and reveals a characteristic nonmonotonic dephasing caused by a sign change in the acceleration-induced frequency difference. For $M=10^6M_\odot$, $m_s/M=10^{-5}$, $a/M=0.7$, and $AM=3\times10^{-7}$, the dominant-mode dephasing slightly exceeds $1$ rad over one year, demonstrating the secular sensitivity of long-duration EMRI waveforms to weak acceleration.

The remainder of this paper is organized as follows. In Sec.~\ref{sec:2}, we formulate the separated Teukolsky equations on the spinning C-metric background and derive their radial source term. In Sec.~\ref{sec:3}, we construct the homogeneous radial solutions and obtain the waveform amplitudes using the Sasaki--Nakamura and Green function methods. In Sec.~\ref{sec:4}, we calculate the energy fluxes, evolve the near-equatorial inspiral, and quantify the acceleration-induced frequency and phase differences. We summarize our results and discuss the limitations of our approximation in Sec.~\ref{sec:5}. The Newman--Penrose quantities, angular solution, near-equatorial circular orbit, source-term coefficients, and details of the Sasaki--Nakamura transformation are collected in the appendices. Unless otherwise stated, we use geometrized units $G=c=1$ and the metric signature $(-,+,+,+)$. In addition, all physical quantities in the numerical calculations of Sec.~\ref{sec:4} are scaled by the black-hole mass $M$.

\section{Teukolsky formalism}
\label{sec:2}
In this section, we formulate the spin-$-2$ Teukolsky equation on the spinning C-metric background, separate it into radial and angular equations, and derive the analytical source term of the radial Teukolsky equation.

\subsection{Teukolsky equation}
\label{sec:2.1}
In the Boyer--Lindquist (BL) coordinates, the spinning C metric describing a rotating accelerating black hole takes the form \cite{Griffiths:2005se}:
\begin{equation}
\label{eq:1}
{\mathrm{d}}{s^2} = {g_{tt}}{\mathrm{d}}{t^2} + 2{g_{t\varphi }}{\mathrm{d}}t{\mathrm{d}}\varphi  + {g_{rr}}{\mathrm{d}}{r^2} + {g_{\theta \theta }}{\mathrm{d}}{\theta ^2} + {g_{\varphi \varphi }}{\mathrm{d}}{\varphi ^2},
\end{equation}
with metric components:
\begin{equation}
\label{eq:2}
\begin{split}
{g_{tt}} &=  - \frac{1}{{{\Omega ^2}}}\frac{1}{\Sigma }\left( {Q - {a^2}P{{\sin }^2}\theta } \right),
\\
{g_{t\varphi }} &= \frac{1}{{{\Omega ^2}}}\frac{{a\,{{\sin }^2}\theta }}{\Sigma }\left[ {Q - P\left( {{r^2} + {a^2}} \right)} \right],
\\
{g_{rr}} &= \frac{1}{{{\Omega ^2}}}\frac{\Sigma }{Q},
\\
{g_{\theta \theta }} &= \frac{1}{{{\Omega ^2}}}\frac{\Sigma }{P},
\\
{g_{\varphi \varphi }} &= \frac{1}{{{\Omega ^2}}}\frac{{{{\sin }^2}\theta }}{\Sigma }\left[ {P{{\left( {{r^2} + {a^2}} \right)}^2} - {a^2}Q{{\sin }^2}\theta } \right],
\end{split}
\end{equation}
where the metric functions $\Omega$, $\Sigma$, $Q$, and $P$ are
\begin{equation}
\label{eq:3}
\begin{split}
\Omega  &= 1 - A r \cos\theta,
\\
\Sigma  &= {r^2} + {a^2}{\cos ^2}\theta,
\\
Q &= \left( {1 - {A^2}{r^2}} \right)\left( {{r^2} - 2Mr + {a^2}} \right),
\\
P &= P(\theta) = 1 - 2 A M  \cos\theta  + {a^2}{A^2}{\cos ^2}\theta.
\end{split}
\end{equation}
The parameters $A$, $a$, and $M$ denote the acceleration, spin, and mass of the black hole, respectively, and the Kerr metric is recovered when $A=0$. For $A\neq0$, one generally has $P(0)\neq P(\pi)$, so the two parts of the symmetry axis cannot both be made regular by a single choice of the azimuthal period. Following \cite{Bini:2008mzd}, we remove the conical singularity at $\theta=\pi$ by taking $0\leq\varphi<2\pi/P(\pi)$. The remaining conical defect at $\theta=0$ represents the axial source that supplies the acceleration---a cosmic string for a deficit, or a strut for an excess. This identification also implies the azimuthal-mode condition $m=m_0P(\pi)$, with $m_0\in\mathbb{Z}$, used below. The geometry contains three horizons: the Cauchy horizon $r_- = M-\sqrt{M^2-a^2}$, the event horizon $r_+ = M+\sqrt{M^2-a^2}$, and the acceleration horizon $r_A=1/A$ \cite{Appels:2016toa}.
	
Unlike Kerr spacetime, the static exterior region of the spinning C metric is bounded by the acceleration horizon and does not approach a standard Kerr-like asymptotically flat wave zone as $r\to\infty$. Consequently, the usual Kerr outgoing-wave behavior and gravitational-wave luminosity formula at null infinity cannot be imported directly. This creates a technical difficulty in defining and extracting the radiative energy flux. For sufficiently small acceleration, we handle this difficulty approximately by placing the observer inside the acceleration horizon and introducing an intermediate extraction radius $r_\infty$ satisfying
\begin{equation}
r_+\ll r_\infty\ll r_A,
\end{equation}
together with $M/r_\infty\ll1$ and $A r_\infty\ll1$. We refer to this intermediate surface as the Kerr-like infinity. There $\Omega=1+\mathcal{O}(A r_\infty)$ and the geometry can be treated as a Kerr-like wave zone, allowing us to use the Kerr outgoing-wave behavior and luminosity formula as approximations. This construction does not assert that the global spacetime is asymptotically flat. Whenever quantities are evaluated or matched at $r_\infty$, we consistently expand them in the small acceleration and retain terms only through first order in $A$, while discarding all terms of $\mathcal{O}(A^2)$ and higher. The wave-zone approximation consequently requires the relevant dimensionless combinations, in particular $AM$, $Aa$, and $A r_\infty$, to remain much smaller than unity.

Away from the conical axis, the spinning C metric is a Ricci-flat vacuum solution and is algebraically special of Petrov type D. For vacuum Petrov type-D backgrounds, the Teukolsky formalism provides decoupled perturbation equations \cite{Teukolsky:1973ha} within the Newman--Penrose formalism \cite{Newman:1961qr}. For the spinning C metric, we first introduce a Newman--Penrose null tetrad $\left( {{l^\mu },{n^\mu },{m^\mu },{{\bar m}^\mu }} \right)$ used by \cite{Chen:2024rov}, and then 12 spin coefficients and five Weyl scalars can be defined by the tetrad and the metric \cite{Compere:2017hsi}. The tetrad and relevant definitions are detailed in Appendix~\ref{A}. The 12 spin coefficients in our convention become
\begin{equation}
\label{eq:4}
\begin{split}
\kappa  &= \lambda  = \sigma  = \nu  = \varepsilon  = 0,
\\
\rho  &= \Omega \left( {1 - i a A {{\cos }^2}\theta } \right){\rho _r},
\\
\mu  &= \frac{Q}{{2{\Omega ^2}\Sigma }}\rho,
\\
\tau & = \frac{{\sqrt P \left( {{r^2}A - ia} \right)\sin \theta }}{{\sqrt 2 \Sigma }},
\\
\varpi  &=  - \frac{{\sqrt P \left( {{r^2}A - ia} \right)\sin \theta }}{{\sqrt 2 }}{\rho _r}^2,
\\
\gamma  &= \mu  + \frac{{{\partial _r}\left( Q \right)\Omega  + 4QA\cos \theta }}{{4\Omega \Sigma }},
\\
\beta  &=  - \frac{{\sqrt P }}{{2\sqrt 2 }}\cot \theta \left( {\frac{{\bar \rho }}{\Omega } + A\cos \theta } \right) + \frac{\Omega }{{2\sqrt 2 }}\frac{{{\partial _\theta }\left( {\sqrt P } \right)}}{{r + ia\cos \theta }},
\\
\alpha  &= \varpi  - \bar \beta  + \frac{{\sqrt {2P} Ar\sin \theta }}{{r - ia\sin \theta }},
\end{split}
\end{equation}
where $\rho_r=-1/(r-ia\cos\theta)$ and the overbar denotes complex conjugation. The 5 Weyl scalars  become
\begin{equation}
\label{eq:5}
\begin{split}
{\psi _0} &= {\psi _1} = {\psi _3} = {\psi _4} = 0,
\\
{\psi _2} &= \left( {1 + iaA} \right)M{\left( {\Omega {\rho _r}} \right)^3}.
\end{split}
\end{equation}
The null tetrad used above is aligned with the two repeated principal null directions, as is also manifest from Eq.~\eqref{eq:5}: only the Coulomb Weyl scalar $\psi_2$ is nonzero in the background. These properties are precisely the local assumptions required for the decoupled Newman--Penrose perturbation equation derived by Teukolsky. We may therefore apply the spin-$-2$ Teukolsky equation in the regular vacuum region of the spacetime of the spinning C metric.

For the spinning C metric, its sourced Teukolsky equation for a gravitational perturbation of spin weight $s=-2$ is
\begin{equation}
\label{eq:6}
\begin{split}
\Big[&(D + 3\gamma - \bar{\gamma} + 4\mu + \bar{\mu})
(D + 4\epsilon - \rho) \\
&- (\bar{\delta} - \bar{\tau} + \bar{\beta} + 3\alpha + 4\varpi)
(\delta - \tau + 4\beta) \\
&- 3\psi_2 \Big] \delta \psi_4
= 4\pi {\mathcal T}_4.
\end{split}
\end{equation}
Here $\mathcal T_4$, derived in Sec.~\ref{sec:2.3}, is the tetrad-projected source. The operators $\mathcal D$, $D$, $\delta$, and $\bar\delta$ denote directional derivatives along the tetrad legs:
\begin{equation}
\label{eq:7}
\begin{aligned}
{\mathcal D} &= {n^\mu }{\nabla _\mu }, &
D &= {l^\mu }{\nabla _\mu },\\
\delta &= {m^\mu }{\nabla _\mu }, &
\bar \delta &= {{\bar m}^\mu }{\nabla _\mu }.
\end{aligned}
\end{equation}
When these operators act on the scalar quantity $\delta\psi_4$, $\nabla_\mu$ reduces to $\partial_\mu$, and the directional derivatives become
\begin{equation}
\label{eq:8}
\begin{split}
{\mathcal D} &= \frac{1}{2\Sigma} \left( (r^2 + a^2) \partial_t - Q \partial_r + a \partial_\varphi \right), 
\\
D &= \Omega^2 \left( \frac{r^2 + a^2}{Q} \partial_t + \partial_r + \frac{a}{Q} \partial_\varphi \right), 
\\
\delta &=\! \frac{\Omega}{\sqrt{2} P (r \!+\! i a \cos\theta)} \left( P \partial_\theta \!+\! \frac{i}{\sin\theta} \partial_\varphi \!+\! i a \sin\theta \partial_t \right), 
\\
\bar{\delta} &=\! \frac{\Omega}{\sqrt{2} P (r \!-\! i a \cos\theta)} \left(P \partial_\theta \!-\! \frac{i}{\sin\theta} \partial_\varphi \!-\!i a \sin\theta  \partial_t \right).
\end{split}
\end{equation}

\subsection{{Separation of variables}}
\label{sec:2.2}
Reference \cite{Bini:2008mzd} showed that the Teukolsky equation for the spinning C metric Eq.~\eqref{eq:6} admits separable solutions of the form:
\begin{equation}
\label{eq:9}
\psi  = {\Omega ^{ - 3}}\frac{1}{{\sqrt {2\pi } }}\int_{ - \infty }^\infty  {{\mathrm{d}}\omega } {\sum\limits_{lm} R\left( r \right)S\left( \theta  \right){e^{im\varphi }} {e^{ - i\omega t}}},
\end{equation}
where $\psi  = {\left[ {\left( {1 - iaA{{\cos }^2}\theta } \right)\Omega {\rho _r}} \right]^{ - 4}}\delta {\psi _4}$. In our paper, we adopt the simpler form $\psi  = {\left( {\Omega {\rho _r}} \right)^{ - 4}}\delta {\psi _4}$ that is used in \cite{Chen:2024rov}. The factor $1/\sqrt{2\pi}$ implements a unitary Fourier transform in time. For the removal of the conical singularity at $\theta  = \pi$, the azimuthal separation constant $m$ should be of the form \cite{Bini:2008mzd}:
\begin{equation}
\label{eq:10}
m = m_0 P(\pi).
\end{equation}

Substituting Eq.~\eqref{eq:9} into Eq.~\eqref{eq:6}, one can get the radial and angular perturbation equations as follows:
\begin{equation}
\label{eq:11}
{Q^2}{\partial _r}\left( {{Q^{ - 1}}{\partial _r}R\left( r \right)} \right) - {V_R}R\left( r \right) = T,
\end{equation}
\begin{equation}
\label{eq:12}
\csc \theta {\partial _\theta }\left( {P\sin \theta {\partial _\theta }S\left( \theta  \right)} \right) + {V_S}S\left( \theta  \right) = 0,
\end{equation}
with effective potentials
\begin{equation}
\label{eq:13}
\begin{split}
V_R=&\; 6rA^2(r-M)
-\frac{[(r^2+a^2)\omega-am]^2}{Q}\\
&\;-4i\biggl[
\frac{\omega M(r^2-a^2)}{r^2-2Mr+a^2}-\frac{\omega r(1+a^2A^2)}{1-A^2r^2}\\
&\;-\frac{am\,\partial_rQ}{2Q}\biggr]+\lambda,
\end{split}
\end{equation}
\begin{equation}
\label{eq:14}
\begin{split}
V_S=&\;-\frac{1}{P\sin^2\theta}
\Big[(a\omega-2AM)\cos^2\theta + m-2AM\\
&\;-a\omega +2(1+a^2A^2)\cos\theta \Big]^2+8m\cot\theta\csc\theta\\
&\;-2A\cos\theta(a^2A\cos\theta-M)-2+2a^2A^2+\lambda.
\end{split}
\end{equation}
Here $\lambda$ is the angular separation constant. For $s=-2$, $A=0$, and $a\omega\ll1$, it reduces to $\lambda=l(l+1)-2+\mathcal{O}(a\omega)$, where $l$ labels the angular mode. At finite $a\omega$ and $A$, we determine $\lambda$ with the continued-fraction method summarized in Appendix~\ref{B} \cite{Chen:2024rov}. The source term of the radial Teukolsky equation is defined as
\begin{equation}
\label{eq:15}
\begin{split}
T=&\;\frac{P(\pi)}{2\pi\sqrt{2\pi}}
\int_{-\infty}^{\infty}\!\mathrm{d}t
\int_{-1}^{1}\!\mathrm{d}(\cos\theta)
\int_0^{2\pi/P(\pi)}\!\mathrm{d}\varphi\\
&\;\times\Omega^{-3}e^{i(\omega t-m\varphi)}
4\pi\Sigma\bigl(-2\mathcal T_4\rho_r^{-4}\bigr)S(\theta),
\end{split}
\end{equation}
where the angular solution is normalized by
\begin{equation}
\label{eq:16}
\int_{ - 1}^1 {\mathrm{d}} \cos \theta {\left( {S\left( \theta  \right)} \right)^2} = 1.
\end{equation}
This convention differs from that of the Black Hole Perturbation Toolkit, whose Kerr angular solutions satisfy the same integral with value $1/(2\pi)$ \cite{BHToolkit}. Reference~\cite{Chen:2024rov} expresses an unnormalized Frobenius-series solution of Eq.~\eqref{eq:12} based on the Heun’s equation. Appendix~\ref{B} explains how we rescale that series to obtain a numerically well-conditioned solution and impose the normalization in Eq.~\eqref{eq:16}.

\subsection{Source term}
\label{sec:2.3}
The tetrad-projected source $\mathcal T_4$ in Eq.~\eqref{eq:6} is related to the stress--energy tensor by
\begin{equation}
\label{eq:17}
\begin{split}
\mathcal T_4 &\;=
(\mathcal D-\bar\gamma+\bar\mu+3\gamma+4\mu)
\Big[(\bar\delta-2\bar\tau+2\alpha)T_{\bar mn}\\
&\;-(\mathcal D+\bar\mu-2\bar\gamma+2\gamma)
T_{\bar m\bar m}\Big]\\
&\;+(\bar\delta+3\alpha+\bar\beta+4\varpi-\bar\tau)
\Big[(\mathcal D+2\bar\mu+2\gamma)T_{\bar mn}\\
&\;-(\bar\delta+2\alpha+2\bar\beta-\bar\tau)T_{nn}\Big],
\end{split}
\end{equation}
where the required tetrad projections of the stress--energy tensor are
\begin{equation}
\label{eq:18}
\begin{aligned}
T_{nn}&=T_{\mu\nu}n^\mu n^\nu,\\
T_{\bar mn}&=T_{\mu\nu}\bar m^\mu n^\nu,\\
T_{\bar m\bar m}&=T_{\mu\nu}\bar m^\mu\bar m^\nu,
\end{aligned}
\end{equation}
and
\begin{equation}
\label{eq:19}
\begin{split}
T^{\mu\nu}=&\;\frac{m_s}{\sqrt{-g}}\frac{u^\mu u^\nu}{u^t}
\delta\bigl(r-r_{\mathrm{ob}}(t)\bigr)\\
&\;\times\delta\bigl(\theta-\theta_{\mathrm{ob}}(t)\bigr)
\delta\bigl(\varphi-\varphi_{\mathrm{ob}}(t)\bigr).
\end{split}
\end{equation}
For the spinning C metric, $\sqrt{-g}=\Sigma\sin\theta/\Omega^4$. Here $m_s$ is the test-particle mass (for EMRI, it equals to the secondary mass), $\delta$ denotes the Dirac delta, the subscript ``$\mathrm{ob}$'' labels the test-particle trajectory, and $u^\mu$ is the test-particle four-velocity satisfying
\begin{equation}
\label{eq:20}
u^\mu = (\dot{t}, \dot{r}, \dot{\theta}, \dot{\varphi}),
\end{equation}
where the dot denotes differentiation with respect to the proper time along the geodesic. In Kerr spacetime, the north--south reflection symmetry makes $\theta=\pi/2$ an invariant geodesic plane, so an orbit initially tangent to this plane remains on it. The spinning C metric does not possess this symmetry: the terms $\Omega$ and $P$ imply $g_{\mu\nu}(r,\pi-\theta)\neq g_{\mu\nu}(r,\theta)$ for $A\neq0$. In particular, $\partial_\theta\Omega$ and $\partial_\theta P$ do not vanish at $\theta=\pi/2$, and the polar geodesic equation is therefore not satisfied there automatically. Consequently, one cannot obtain a circular timelike orbit simply by imposing the Kerr equatorial condition, and the orbital polar position must be solved together with its energy and angular momentum.

To describe the orbital radial--polar stability, we introduce the two-dimensional effective potential \cite{Tahara:2024pot}:
\begin{equation}
\label{effective potential}
\begin{split}
V_{\mathrm{eff}}(r,\theta;\mathcal{E},\mathcal{L})
={}&1+g_{tt}\dot t^2+2g_{t\varphi}\dot t\dot\varphi
+g_{\varphi\varphi}\dot\varphi^2
\\
={}&-g_{rr}\dot r^2-g_{\theta\theta}\dot\theta^2.
\end{split}
\end{equation}
Stationarity and axisymmetry of the spinning C metric provide the conserved specific energy $\mathcal{E}$ and specific angular momentum $\mathcal{L}$, which determine the above $\dot t$ and $\dot\varphi$. For a circular orbit at fixed $(r_{\mathrm{ob}},\theta_{\mathrm{ob}})$, one has $\dot r=\dot\theta=0$, and the effective potential should satisfy
\begin{equation}
\label{circular orbit conditions}
V_{\mathrm{eff}}(r_{\mathrm{ob}},\! \theta_{\mathrm{ob}})
\!=\! \partial_r\! V_{\mathrm{eff}}(r_{\mathrm{ob}} ,\!  \theta_{\mathrm{ob}})
\!=\! \partial_\theta\! V_{\mathrm{eff}}(r_{\mathrm{ob}} ,\! \theta_{\mathrm{ob}})\!=\!0.
\end{equation}
The above three conditions determine $\mathcal{E}$, $\mathcal{L}$, and $\theta_{\mathrm{ob}}$ for a specified radius $r_{\mathrm{ob}}$. We seek their regular continuation at small acceleration in the form
\begin{equation}
\label{small acceleration form}
\begin{split}
\theta_{\mathrm{ob}} =&\; \frac{\pi}{2}+A f_1(r_{\mathrm{ob}})+A^2f_2(r_{\mathrm{ob}})+\mathcal{O}(A^3),\\
\mathcal{E} =&\; \mathcal{E}_{\mathrm{K}} + A\mathcal{E}_1 + A^2\mathcal{E}_2 + \mathcal{O}(A^3),\\
\mathcal{L}=&\; \mathcal{L}_{\mathrm{K}} + A\mathcal{L}_1 + A^2\mathcal{L}_2 + \mathcal{O}(A^3),
\end{split}
\end{equation}
where $\mathcal{E}_{\mathrm{K}}$ and $\mathcal{L}_{\mathrm{K}}$ represent Kerr results, and all the coefficients are independent of $A$. Expanding the conditions in Eq.~\eqref{circular orbit conditions} through first order in $A$, we have $\mathcal{E}_1=\mathcal{L}_1=0$ and
\begin{equation}
\label{small acceleration results 1}
f_1(r_{\mathrm{ob}})=\frac{f_N(r_{\mathrm{ob}})}{f_D(r_{\mathrm{ob}})},
\end{equation}
with
\begin{equation}
\label{small acceleration results 2}
\begin{split}
f_N(r_{\mathrm{ob}})=&\; r_{\mathrm{ob}}^4 - 3M r_{\mathrm{ob}}^3 + 2a M^{1/2} r_{\mathrm{ob}}^{5/2} + M^2 r_{\mathrm{ob}}^2\\
&\;- 2a M^{3/2} r_{\mathrm{ob}}^{3/2} + a^2 M r_{\mathrm{ob}},\\
f_D(r_{\mathrm{ob}})=&\;M r_{\mathrm{ob}}^2 - 4a M^{3/2} r_{\mathrm{ob}}^{1/2} + 3a^2 M.
\end{split}
\end{equation}
 The orbit is displaced from the Kerr equatorial plane by an angle of order \(A\), and we therefore refer to this orbit as the near-equatorial circular orbit. The orbital stability is tested by the metric-weighted Hessian of $V_{\mathrm{eff}}$, and more details are shown in Appendix~\ref{C}.

For later convenience, we write the three tetrad projections as
\begin{equation}
\label{eq:21}
\begin{split}
T_{nn}=&\; m_s\frac{\Omega^4C_{nn}}{\sin\theta}
\delta\bigl(r-r_{\mathrm{ob}}(t)\bigr)\\[-2pt]
&\;\times\delta\bigl(\theta-\theta_{\mathrm{ob}}(t)\bigr)
\delta\bigl(\varphi-\varphi_{\mathrm{ob}}(t)\bigr),\\
T_{\bar mn}=&\; m_s\frac{\Omega^4C_{\bar mn}}{\sin\theta}
\delta\bigl(r-r_{\mathrm{ob}}(t)\bigr)\\[-2pt]
&\;\times\delta\bigl(\theta-\theta_{\mathrm{ob}}(t)\bigr)
\delta\bigl(\varphi-\varphi_{\mathrm{ob}}(t)\bigr),\\
T_{\bar m\bar m}=&\; m_s\frac{\Omega^4C_{\bar m\bar m}}{\sin\theta}
\delta\bigl(r-r_{\mathrm{ob}}(t)\bigr)\\[-2pt]
&\;\times\delta\bigl(\theta-\theta_{\mathrm{ob}}(t)\bigr)
\delta\bigl(\varphi-\varphi_{\mathrm{ob}}(t)\bigr),
\end{split}
\end{equation}
where
\begin{equation}
\label{eq:22}
\begin{split}
{C_{nn}} &= \frac{1}{{\Sigma {u^t}}}{g_{\mu \alpha }}{g_{\nu \beta }}{u^\alpha }{u^\beta }{n^\mu }{n^\nu },
\\
{C_{\bar mn}} &= \frac{1}{{\Sigma {u^t}}}{g_{\mu \alpha }}{g_{\nu \beta }}{u^\alpha }{u^\beta }{{\bar m}^\mu }{n^\nu },
\\
{C_{\bar m\bar m}} &= \frac{1}{{\Sigma {u^t}}}{g_{\mu \alpha }}{g_{\nu \beta }}{u^\alpha }{u^\beta }{{\bar m}^\mu }{{\bar m}^\nu }.
\end{split}
\end{equation}
By using Eqs.~\eqref{eq:17} and \eqref{eq:21}, it is convenient to rewrite the radial source term Eq.~\eqref{eq:15} further as
\begin{equation}
\label{eq:23}
\begin{split}
T &\;=  m_s \int_{-\infty}^{\infty} \mathrm{d}t \, e^{i(\omega t - m\varphi)} Q^2 \\
&\;\times \Big[ 
( A_{nn0} + A_{\bar mn0}+ A_{\bar m\bar m0} ) \delta \left( r - r_{\mathrm{ob}}(t) \right) \\
&\;+ \partial_r \bigl( \left( A_{\bar mn1} + A_{\bar m\bar m1} \right) \delta \left( r - r_{\mathrm{ob}}(t) \right) \bigr)\\
&\;+ \partial_r \partial_r \bigl( A_{\bar m\bar m2} \delta \left( r - r_{\mathrm{ob}}(t) \right) \bigr) 
\Big] \Bigg|_{\theta = \theta_{\mathrm{ob}}(t),\; \varphi = \varphi_{\mathrm{ob}}(t)},
\end{split}
\end{equation}
where the source-term coefficients $A_{nn0}$, $A_{\bar mn0}$, $A_{\bar m\bar m0}$, $A_{\bar mn1}$, $A_{\bar m\bar m1}$, and $A_{\bar m\bar m2}$ are given in Appendix~\ref{D}, and their limits of $A\to0$ agree with the standard Kerr source-term coefficients \cite{Mino:1997bx,Hughes:1999bq,Sasaki:2003xr}. By invoking these coefficients, the subsequent waveform amplitudes can be computed simply.

\section{Waveform amplitudes}
\label{sec:3}
This section constructs the radial waveform amplitudes in three steps. We first transform the homogeneous Teukolsky equation to Sasaki--Nakamura (SN) form and derive relevant ingoing and outgoing boundary conditions. We then integrate the two independent ingoing and outgoing solutions of the SN equation from their respective boundaries to the orbital radius. Finally, the waveform amplitudes at the event horizon and at the Kerr-like extraction surface are then obtained by combining these solutions with the radial source term via a Green-function construction.

\subsection{Sasaki-Nakamura transformation}
\label{sec:3.1}
We begin with the homogeneous part of the radial Teukolsky equation in Eq.~\eqref{eq:11}:
\begin{equation}
\label{eq:24}
{Q^2}{\partial _r}\left( {{Q^{ - 1}}{\partial _r}R\left( r \right)} \right) - {V_R}R\left( r \right) = 0.
\end{equation}
The Teukolsky radial potential is long ranged, which poses a challenge for numerical integrations \cite{Lo:2023fvv}. The SN transformation converts the homogeneous equation into one with a short-range potential and well-controlled wave behaviors \cite{Sasaki:1981kj,Sasaki:1981sx}, so that accurate and efficient numerical integrations can be carried out. The source term Eq.~\eqref{eq:23} is reintroduced later through the Green function \cite{Mino:1997bx,Hughes:1999bq,Sasaki:2003xr}. Applying the SN transformation to Eq.~\eqref{eq:24} gives its SN equation:
\begin{equation}
\label{eq:25}
{\partial _{{r_*}}}\left( {{\partial _{{r_*}}}X\left( r \right)} \right) - {F_1}{\partial _{{r_*}}}X\left( r \right) - {U_1}X\left( r \right) = 0,
\end{equation}
where $F_1$ and $U_1$ are defined in Appendix~\ref{E}, and
\begin{equation}
\label{eq:26}
{\partial _{{r_*}}} = \frac{Q}{{{r^2} + {a^2}}}{\partial _r}.
\end{equation}
The tortoise coordinate $r_*$ is obtained by integrating Eq.~\eqref{eq:26}, and its complete expression is shown in Appendix~\ref{E}. Defining $\Delta=r^2-2Mr+a^2$, one has $Q=\Delta+\mathcal{O}(A^2)$. Hence there is no correction linear in $A$, and
\begin{equation}
\label{eq:27}
r_*=r_*^{\mathrm K}+\mathcal{O}(A^2),
\end{equation}
where the integration constant chosen to reproduce the standard Kerr convention, and the tortoise coordinate for the Kerr case is
\begin{equation}
\label{eq:28}
\begin{split}
r_*^{\mathrm K}=&\; r+\frac{2Mr_+}{r_+-r_-}
\ln\left(\frac{r-r_+}{2M}\right)\\
&\; -\frac{2Mr_-}{r_+-r_-}
\ln\left(\frac{r-r_-}{2M}\right),
\end{split}
\end{equation}
Thus the exact but lengthy acceleration-dependent form for $r_*$ is unnecessary in a calculation consistently truncated at first order in $A$. The forward transformation from $R(r)$ to $X(r)$ is
\begin{equation}
\label{eq:29}
X(r) = \sqrt{Q^{-2} (r^2 + a^2)} \left( \alpha R(r) + \beta Q^{-1} \partial_r R(r) \right),
\end{equation}
and the inverse transformation from $X(r)$ to $R(r)$ is
\begin{equation}
\label{eq:30}
\begin{split}
R(r) =&\; \frac{1}{\eta} \Bigg[
\big( \alpha + Q^{-1} \partial_r \beta \big) 
\frac{X(r)}{\sqrt{Q^{-2} (r^2 + a^2)}} \\
&\;- \beta Q^{-1} \partial_r \bigg( \frac{X(r)}{\sqrt{Q^{-2} (r^2 + a^2)}} \bigg)
\Bigg],
\end{split}
\end{equation}
where the unexpanded definitions of $\alpha$, $\beta$, and $\eta$ are collected in Appendix~\ref{E}.

At the event horizon and the Kerr-like extraction surface, the first-order limiting values of $F_1$ and $U_1$ are
\begin{equation}
\label{eq:31}
\begin{split}
F_1(r\to r_+)&=0,\\
F_1(r\to r_\infty)&=\mathcal O(r_\infty^{-2})
+\mathcal O(A^2),\\
U_1(r\to r_+)&=-k^2,\\
U_1(r\to r_\infty)&=-\omega^2+\mathcal O(r_\infty^{-2})
+\mathcal O(A^2),
\end{split}
\end{equation}
where
\begin{equation}
\label{eq:32}
k = \omega - \frac{m a}{2 M r_+}.
\end{equation}
These limiting values define the homogeneous solution $X^{\mathrm{in}}$, which is purely ingoing at the event horizon, and the solution $X^{\mathrm{up}}$, which is purely outgoing at the Kerr-like extraction surface. Their boundary behaviors satisfy
\begin{equation}
\label{eq:33}
X^{\mathrm{in}} \to 
\begin{cases}
\,B_{\mathrm{SN}}^{\mathrm{trans}} e^{-ikr_*}, & r \to r_+ 
\\[8pt]
\,B_{\mathrm{SN}}^{\mathrm{ref}} e^{i\omega r_*} + B_{\mathrm{SN}}^{\mathrm{inc}} e^{-i\omega r_*}, & r \to r_\infty
\end{cases},
\end{equation}
\begin{equation}
\label{eq:34}
X^{\mathrm{up}} \to 
\begin{cases}
\,C_{\mathrm{SN}}^{\mathrm{inc}} e^{ikr_*} + C_{\mathrm{SN}}^{\mathrm{ref}} e^{-ikr_*}, \; & r \to r_+ 
\\[8pt]
\,C_{\mathrm{SN}}^{\mathrm{trans}} e^{i\omega r_*}, \; & r \to r_\infty
\end{cases}.\,
\end{equation}
The coefficients multiplying the waves are independent of $r$. We choose a normalization convention:
\begin{equation}
\label{eq:35}
B_{\mathrm{SN}}^{\mathrm{trans}} = C_{\mathrm{SN}}^{\mathrm{trans}} = 1.
\end{equation}
The transformation relations Eqs.~\eqref{eq:29} and \eqref{eq:30} then give the corresponding boundary behaviors of the Teukolsky function $R\left( r \right)$:
\begin{equation}
\label{eq:36}
R^{\mathrm{in}} \to 
\begin{cases}
\,B_{\mathrm{TK}}^{\mathrm{trans}} Q^2 e^{-ikr_*}, & r \to r_+ 
\\[8pt]
\,B_{\mathrm{TK}}^{\mathrm{ref}} r^3 e^{i\omega r_*} + B_{\mathrm{TK}}^{\mathrm{inc}} r^{-1} e^{-i\omega r_*}, & r \to r_\infty
\end{cases},
\end{equation}
\begin{equation}
\label{eq:37}
R^{\mathrm{up}} \to 
\begin{cases}
\,C_{\mathrm{TK}}^{\mathrm{inc}} e^{ikr_*} + C_{\mathrm{TK}}^{\mathrm{ref}} Q^2 e^{-ikr_*}, \quad\;\, & r \to r_+ 
\\[8pt]
\,C_{\mathrm{TK}}^{\mathrm{trans}} r^3 e^{i\omega r_*}, \quad\;\, & r \to r_\infty\,
\end{cases}.
\end{equation}
The above equations reduce to the Kerr asymptotic behaviors when $A=0$. Under the SN transformation, the transmitted Teukolsky and SN amplitudes are related by constant conversion factors \cite{Lo:2023fvv}. Adopting the normalization convention Eq.~\eqref{eq:35}, we obtain
\begin{equation}
\label{eq:38}
B_{\mathrm{TK}}^{\mathrm{trans}} = \frac{B_{\mathrm{TK}}^{\mathrm{trans}}}{B_{\mathrm{SN}}^{\mathrm{trans}}} = \frac{1}{s_1},
\end{equation}
\begin{equation}
\label{eq:39}
C_{\mathrm{TK}}^{\mathrm{trans}} = \frac{C_{\mathrm{TK}}^{\mathrm{trans}}}{C_{\mathrm{SN}}^{\mathrm{trans}}} = {s_2},
\end{equation}
with
\begin{equation}
\label{eq:40}
\begin{aligned}
s_1 =&\;4\sqrt{2 M r_+} \big[ (2 - 6 i M \omega - 4 M^2 \omega^2) r_+^2 \\
&\;+ (3 i a m - 4 M + 4 a m M \omega + 6 i M^2 \omega) r_+ \\
&\;- a^2 m^2 - 3 i a m M + 2 M^2 \big] + \mathcal{O}(A^2),
\\
s_2 = &-\frac{4\omega^2}{2\lambda + \lambda^2 - 12\omega(iM + a^2\omega - am)},
\end{aligned}
\end{equation}
where $m$, $\omega$, and $\lambda$ retain their implicit dependence on $A$. When $A=0$, $s_1$ and $s_2$  reduce directly to the standard Kerr results \cite{Lo:2023fvv}. Due to the intermediate-wave-zone approximation at $r_\infty$, $s_2$ takes the same form as the Kerr conversion factor.

\subsection{Numerical integration}\label{sec:3.2}
Using Eq.~\eqref{eq:26} to replace $r_*$ derivatives by $r$ derivatives, the radial perturbation equation Eq.~\eqref{eq:25} becomes
\begin{equation}
\label{eq:41}
\partial_r \partial_r X(r) + p \partial_r X(r) + q X(r) = 0,
\end{equation}
where
\begin{equation}
\label{eq:42}
\begin{split}
p&=-\frac{(a^2+r^2)^2F_1+2rQ-(a^2+r^2)\partial_rQ}
{(a^2+r^2)Q},\\
q&=-\frac{(a^2+r^2)^2U_1}{Q^2},
\end{split}
\end{equation}
leading to an identical relation:
\begin{equation}
\label{eq:43}
\partial_r \partial_r X(r) = -p \partial_r X(r) - q X(r).
\end{equation}
In order to obtain accurate numerical solutions for the ingoing and outgoing components and avoid the divergence of $p$ and $q$ at the event horizon, we need to include higher-order correction terms in the asymptotic solutions. For $r\to r_+$, the ingoing solution has a formal series expansion of the form:
\begin{equation}
\label{eq:44}
\begin{split}
&\mathcal{X}^{\mathrm{in}}(r)=e^{-ikr_*}\Big[1+\mathcal A^{\mathrm{in}}r_h+\mathcal B^{\mathrm{in}}r_h^2
+\mathcal C^{\mathrm{in}}r_h^3\Big],
\end{split}
\end{equation}
with $r_h = r-r_+$. For $r\to r_\infty$, the outgoing solution has a formal series expansion of the form:
\begin{equation}
\label{eq:45}
\begin{split}
&\mathcal{X}^{\mathrm{up}}(r)=e^{i\omega r_*}\biggl[1+\frac{\mathcal A^{\mathrm{up}}}{\omega r}
+\frac{\mathcal B^{\mathrm{up}}}{(\omega r)^2}
+\frac{\mathcal C^{\mathrm{up}}}{(\omega r)^3}\biggr].
\end{split}
\end{equation}
The coefficients $\mathcal A^{\mathrm{in/up}}$, $\mathcal B^{\mathrm{in/up}}$, and $\mathcal C^{\mathrm{in/up}}$ follow by substituting Eqs.~\eqref{eq:44} and \eqref{eq:45} into Eq.~\eqref{eq:41} and canceling the three leading powers in each boundary expansion \cite{Lo:2023fvv}. Their explicit expressions are lengthy and are not needed for the subsequent discussion.

Based on Eqs.~\eqref{eq:43}, \eqref{eq:44} and \eqref{eq:45}, the homogeneous radial solutions and their derivatives at any position can be obtained by solving the following initial-value problem:
\begin{equation}
\label{eq:46}
\begin{cases}
\,\partial_r \partial_r X^{\mathrm{in/up}}(r) = \!-\!p \partial_r X^{\mathrm{in/up}}(r) \!-\! q X^{\mathrm{in/up}}(r)
\\[8pt]
\,X^{\mathrm{in/up}} \left( r_0^{\mathrm{in/up}} \right)=\mathcal{X}^{\mathrm{in/up}} \left( r_0^{\mathrm{in/up}} \right)
\\[8pt]
\,\partial_r X^{\mathrm{in/up}} \left( r_0^{\mathrm{in/up}} \right)=\partial_r \mathcal{X}^{\mathrm{in/up}} \left( r_0^{\mathrm{in/up}} \right)
\end{cases},
\end{equation}
where the ingoing solution is obtained by an outward integration from the event horizon, whereas the outgoing solution is obtained by an inward integration from the extraction surface. The two integration paths are
\begin{equation}
\label{eq:48}
\begin{cases}
\,r_0^{\mathrm{in}} \to r_{\mathrm{ob}}(t) 
\\[8pt]
\,r_0^{\mathrm{up}} \to r_{\mathrm{ob}}(t)
\end{cases},
\end{equation}
where the numerical starting radii are
\begin{equation}
\begin{cases}
\label{eq:49}
\,r_0^{\mathrm{in}} = r_+ + 10^{-5}M > r_+
\\[8pt]
\,r_0^{\mathrm{up}} = r_\infty=1000M \ll r_A
\end{cases}.
\end{equation}
We set the outer starting radius to \(r_0^{\mathrm{up}}=r_\infty=1000M\). Although the asymptotic expansion in Eq.~(\ref{eq:45}) is truncated at \(\mathcal{O}[(\omega r)^{-3}]\), we have verified its accuracy by repeating the calculation with several larger values of \(r_\infty\). The resulting homogeneous solutions and energy fluxes change only negligibly, indicating that \(r_\infty=1000M\) is sufficiently large for the parameter range considered here. The initial-value problem Eq.~\eqref{eq:46} is solved by using Mathematica's \texttt{NDSolve} which controls the step size of numerical integration adaptively. Both integrations terminate at the instantaneous orbital radius $r_{\mathrm{ob}}(t)$ for the subsequent substitution of the source term Eq.~\eqref{eq:23}. The final numerical data obtained are therefore
\begin{equation}
\label{eq:50}
\left. \left( X^{\mathrm{in/up}}, \partial_r X^{\mathrm{in/up}}, \partial_r \partial_r X^{\mathrm{in/up}} \right) \right|_{\,r = r_{\mathrm{ob}}(t) }.
\end{equation}
The inverse SN transformation in Eq.~\eqref{eq:30} converts these data to
\begin{equation}
\label{eq:51}
\begin{pmatrix}
\begin{split}
&\;X^{\mathrm{in/up}}\;\\
&\;\partial_rX^{\mathrm{in/up}}\;\\
&\;\partial_r^2X^{\mathrm{in/up}}\;
\end{split}
\end{pmatrix}
\xLongrightarrow{\;\mathrm{SN}\;}
\left.
\begin{pmatrix}
\begin{split}
&\;R^{\mathrm{in/up}}\;\\
&\;\partial_rR^{\mathrm{in/up}}\;\\
&\;\partial_r^2R^{\mathrm{in/up}}\;
\end{split}
\end{pmatrix}
\right|_{r=r_{\mathrm{ob}}(t)}.
\end{equation}
Both $X^{\mathrm{in/up}}$ and $R^{\mathrm{in/up}}$ in Eq.~\eqref{eq:51} are homogeneous radial solutions, and the radial source term enters only in the Green-function construction below.

\subsection{Green-function construction}
\label{sec:3.3}
The retarded solution of the inhomogeneous radial equation Eq.~\eqref{eq:11} is constructed from the homogeneous radial solutions $R^{\mathrm{in}}$ and $R^{\mathrm{up}}$, together with the source term $T$ as \cite{Mino:1997bx,Hughes:1999bq,Sasaki:2003xr}
\begin{equation}
\label{eq:52}
\begin{split}
R(r)=&\;\frac{1}{W}\Bigg[R^{\mathrm{up}}(r)
\int_{r_+}^{r}\!\mathrm{d}r'\,
\frac{R^{\mathrm{in}}(r')T(r')}{Q(r')^2}\\
&\; +R^{\mathrm{in}}(r)
\int_{r}^{r_\infty}\!\mathrm{d}r'\,
\frac{R^{\mathrm{up}}(r')T(r')}{Q(r')^2}\Bigg],
\end{split}
\end{equation}
where the $r$-independent Wronskian is given by
\begin{equation}
\label{eq:53}
W = Q(r)^{-1} \! \left( R^{\mathrm{in}}(r) \, \partial_r R^{\mathrm{up}}(r) - R^{\mathrm{up}}(r) \, \partial_r R^{\mathrm{in}}(r) \right).
\end{equation}
The above equation is equal to a constant and can be evaluated at any radius. At the Kerr-like extraction surface $r=r_\infty$, the first-order asymptotics in Eqs.~\eqref{eq:36} and \eqref{eq:37} give
\begin{equation}
\label{eq:54}
\begin{split}
W=&\;\frac{2C_{\mathrm{TK}}^{\mathrm{trans}}
B_{\mathrm{TK}}^{\mathrm{inc}}}
{(r_\infty^2-2Mr_\infty+a^2)^2}\bigl(2a^2r_\infty-4Mr_\infty^2\\
&\;+2r_\infty^3+ia^2r_\infty^2\omega+ir_\infty^4\omega\bigr) +\mathcal O(A^2).
\end{split}
\end{equation}
Taking the additional wave-zone limit $r_\infty/M\gg1$ reduces this expression to
\begin{equation}
\label{eq:55}
W =2 i \omega C_{\mathrm{TK}}^{\mathrm{trans}} B_{\mathrm{TK}}^{\mathrm{inc}} .
\end{equation}
Combining Eqs.~\eqref{eq:53} and \eqref{eq:55} yields
\begin{equation}
\label{eq:56}
B_{\mathrm{TK}}^{\mathrm{inc}}=
\frac{R^{\mathrm{in}}(r)\partial_rR^{\mathrm{up}}(r)
-R^{\mathrm{up}}(r)\partial_rR^{\mathrm{in}}(r)}{2i\omega Q(r)C_{\mathrm{TK}}^{\mathrm{trans}}}.
\end{equation}
For $r=r_{\mathrm{ob}}(t)$, $B_{\mathrm{TK}}^{\mathrm{inc}}$ is computed by Eq.~\eqref{eq:56} and the homogeneous data in Eq.~\eqref{eq:51}.

The inhomogeneous radial solution Eq.~\eqref{eq:52} consequently has the following asymptotic behaviors:
\begin{equation}
\label{eq:57}
\begin{split}
R(r\to r_+)=&\;\tilde{Z}^{\mathrm{up}}Q(r)^2e^{-ikr_*},
\end{split}
\end{equation}
\begin{equation}
\label{eq:58}
\begin{split}
R(r\to r_\infty)=&\;\tilde{Z}^{\mathrm{in}} r^3e^{i\omega r_*},
\end{split}
\end{equation}
with
\begin{equation}
\label{eq:59}
\tilde{Z}^{\mathrm{up}} = \frac{B_{\mathrm{TK}}^{\mathrm{trans}}}{2i \omega C_{\mathrm{TK}}^{\mathrm{trans}} B_{\mathrm{TK}}^{\mathrm{inc}}}
\int_{r_+}^{r_\infty} \mathrm{d}r'  \frac{R^{\mathrm{up}}(r') T(r')}{Q(r')^2},
\end{equation}
\begin{equation}
\label{eq:60}
\tilde{Z}^{\mathrm{in}} = \frac{1}{2i\omega B_{\mathrm{TK}}^{\mathrm{inc}}} \int_{r_+}^{r_\infty} \mathrm{d}r'  \frac{R^{\mathrm{in}}(r') T(r')}{Q(r')^2} .\qquad\;\,
\end{equation}
For circular orbits, the frequency spectrum of the source is both monochromatic and discrete. Accordingly, substituting Eq.~\eqref{eq:23} into Eqs.~\eqref{eq:59} and \eqref{eq:60} gives
\begin{equation}
\label{eq:61}
\tilde{Z}^{\mathrm{up}} = \delta( \omega - m \Omega_{\varphi} ) \, Z^{\mathrm{up}},
\end{equation}
\begin{equation}
\label{eq:62}
\tilde{Z}^{\mathrm{in}} = \delta( \omega - m \Omega_{\varphi} ) \, Z^{\mathrm{in}},
\end{equation}
with
\begin{equation}
\label{eq:63}
\begin{split}
Z^{\mathrm{up}} = &\;m_s
\frac{ B_{\mathrm{TK}}^{\mathrm{trans}} }{ 2 i \omega C_{\mathrm{TK}}^{\mathrm{trans}} B_{\mathrm{TK}}^{\mathrm{inc}} } \\
&\times\Big[ 
\left( A_{nn0} + A_{\bar{m} n0} + A_{\bar{m} \bar{m} 0} \right)
R^{\mathrm{up}}\\
&- \left( A_{\bar{m} n1} + A_{\bar{m} \bar{m} 1} \right)
\partial_r R^{\mathrm{up}}\\
&+A_{\bar{m} \bar{m} 2}\,\partial_r^2R^{\mathrm{up}}
\Big] 
\bigg|_{\substack{r=r_{\mathrm{ob}}(t),\; \theta=\theta_{\mathrm{ob}}(t)}},
\end{split}
\end{equation}
\begin{equation}
\label{eq:64}
\begin{split}
Z^{\mathrm{in}} = &\;m_s \frac{ 1 }{ 2 i \omega B_{\mathrm{TK}}^{\mathrm{inc}} } \\
&\times\Big[ 
\left( A_{nn0} + A_{\bar{m} n0} + A_{\bar{m} \bar{m} 0} \right)
R^{\mathrm{in}}\\
&- \left( A_{\bar{m} n1} + A_{\bar{m} \bar{m} 1} \right)
\partial_r R^{\mathrm{in}}\\
&+A_{\bar{m} \bar{m} 2}\,\partial_r^2R^{\mathrm{in}}
\Big] 
\bigg|_{\substack{r=r_{\mathrm{ob}}(t),\; \theta=\theta_{\mathrm{ob}}(t)}},
\end{split}
\end{equation}
where $Z^{\mathrm{up}}$ is the horizon waveform amplitude, while $Z^{\mathrm{in}}$ is the far-zone waveform amplitude. Here $B_{\mathrm{TK}}^{\mathrm{trans}}$ and $C_{\mathrm{TK}}^{\mathrm{trans}}$ are fixed by Eqs.~\eqref{eq:38} and \eqref{eq:39}; the homogeneous solutions $R^{\mathrm{in/up}}$ and their derivatives at $r = r_{\mathrm{ob}}(t)$ are obtained from Eq.~\eqref{eq:51}; $B_{\mathrm{TK}}^{\mathrm{inc}}$ follows from Eq.~\eqref{eq:56}; the orbital frequency $\omega$ is given in Appendix~\ref{C}; and the source-term coefficients $A_{nn0}$, $A_{\bar mn0}$, $A_{\bar m\bar m0}$, $A_{\bar mn1}$, $A_{\bar m\bar m1}$ and $A_{\bar m\bar m2}$ are listed in Appendix~\ref{D}.

\section{Orbital evolution and dephasing}
\label{sec:4}

In this section, we use the energy luminosities (from now on luminosities are also called “fluxes”, with a slight abuse of terminology) to evolve the orbital radius and integrate the resulting frequency difference to obtain the gravitational-wave dephasing. The following numerical calculations are performed in geometrized units. Physical quantities quoted without explicit units have been scaled by the black-hole mass $M$ so as to be dimensionless. For spin, length, time, acceleration and angular velocity, their dimensionless forms correspond to $a/M$, $r/M$, $t/M$, $AM$ and $\omega M$, respectively. To convert these dimensionless quantities to SI units, we simply multiply them by the corresponding conversion factors: $GM/c$, $GM/c^2$, $GM/c^3$, $c^4/(GM)$ and $c^3/(GM)$, in the same order. For the black hole mass, we take $M = 10^{6}  M_{\odot}$, where $M_{\odot}$ stands for the solar mass. Therefore, these factors are $4.43 \times 10^{17} \, \mathrm{m}^{2}/\mathrm{s}$, $1.477 \times 10^9 \, \mathrm{m}$, $4.926 \, \mathrm{s}$, $6.086 \times 10^{7} \, \mathrm{m}/\mathrm{s}^{2}$ and $0.203 \, \mathrm{rad} / \mathrm{s}$, respectively.

\subsection{Energy fluxes}
\label{sec:4.1}

Because the waveform is extracted in the intermediate region $r_+\ll r_\infty\ll r_A$, where $M/r_\infty\ll1$ and $Ar_\infty\ll1$, we evaluate the outward energy flux at $r_\infty$ with the leading Kerr-form expression \cite{Hughes:1999bq,Piovano:2020zin}:
\begin{equation}
\label{eq:65}
\left\langle \frac{\mathrm{d}E^{\mathrm{up}}}{\mathrm{d}t} \right\rangle
= \sum_{l=2}^{\infty} \sum_{\substack{m_0 \neq 0 \\ m_0=-l}}^{l} \frac{1}{4\pi \omega^2} \left| Z^{\mathrm{in}} \right|^2.\quad\;
\end{equation}
For the parameter range considered below, the magnitude of the horizon flux is typically $10^{-3}$--$10^{-2}$ of the outward flux. Acceleration corrections to the horizon weighting factor consequently have a subleading effect at the present level of accuracy, and we approximate the horizon contribution by its Kerr-form expression:
\begin{equation}
\label{eq:66}
\left\langle \frac{\mathrm{d}E^{\mathrm{in}}}{\mathrm{d}t} \right\rangle = 
\sum_{l=2}^{\infty} 
\sum_{\substack{m_0 \neq 0 \\ m_0 = -l}}^{l} 
\frac{1}{4\pi \omega^2} \alpha_m \left| Z^{\mathrm{up}} \right|^2,
\end{equation}
where
\begin{equation}
\label{eq:67}
\alpha_m = \frac{256 (2Mr_+)^5 \kappa (\kappa^2 + 4\epsilon^2) (\kappa^2 + 16\epsilon^2) \omega^3}{C_m},
\end{equation}
with
\begin{equation}
\label{eq:68}
\begin{split}
\kappa={}&\omega-\frac{ma}{2Mr_+},\\
\epsilon={}&\frac{\sqrt{M^2-a^2}}{4Mr_+},\\
C_m={}&\bigl[(\lambda+2)^2+4am\omega-4a^2\omega^2\bigr]\\
&\times\bigl(\lambda^2+36ma\omega-36a^2\omega^2\bigr)\\
&+(2\lambda+3)\bigl(96a^2\omega^2-48ma\omega\bigr)\\
&+144\omega^2(M^2-a^2).
\end{split}
\end{equation}
The secondary follows the near-equatorial circular orbit constructed in Appendix~\ref{C}. Its radial and polar frequencies vanish, so the general harmonic relation reduces to $\omega=m\Omega_\varphi+k\Omega_\theta+n\Omega_r=m\Omega_\varphi$, with $k=n=0$. The sums in Eqs.~\eqref{eq:65} and \eqref{eq:66} therefore contain only the azimuthal harmonics $m=m_0P(\pi)$. The angular index $l$ is fixed by the branch of the separation eigenvalue $\lambda$. Since the quadrupolar $l=2$ modes dominate the energy and angular-momentum losses for the circular inspirals of interest \cite{Estelles:2020osj}, we retain only $l=2$ in the following orbital evolution. This is a mode truncation, not an exact selection rule.

\begin{figure}[htbp]
\centering
\includegraphics[width=\columnwidth]{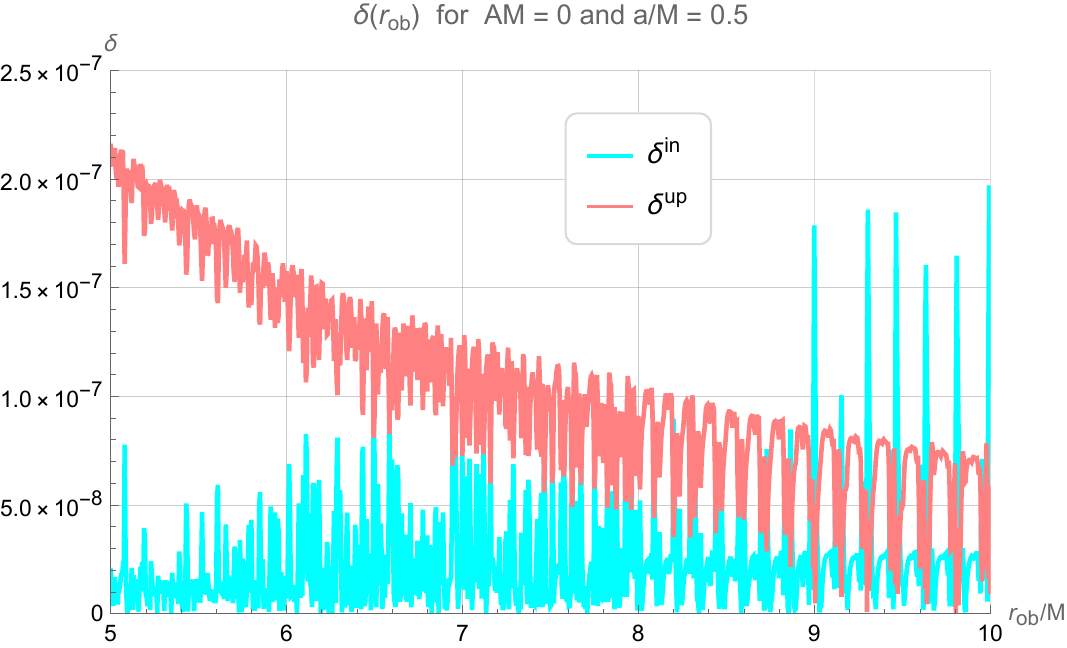}
\caption{\label{fig:1}Relative differences between the SN and MST fluxes, defined in Eq.~\eqref{eq:69}, as functions of the orbital radius. The cyan and pink curves correspond to the event-horizon and outward fluxes, respectively. We set $AM=0$, $a/M=0.5$, $s=-2$, $l=m_0=2$, and use the prograde equatorial circular orbit. The normalization $m_s/M=1$ is used because the plotted relative differences are independent of the overall mass-ratio scaling.}
\end{figure}

As a validation of the radial solver and normalization, we set $A=0$ and compare the event-horizon and outward fluxes obtained with our SN implementation against independent Mano--Suzuki--Takasugi (MST) results \cite{Mano:1996vt} from the Black Hole Perturbation Toolkit \cite{BHToolkit}. We define the relative differences by
\begin{equation}
\label{eq:69}
\delta^{\mathrm{(in/up)}}(r_{\mathrm{ob}}) = 
\left|
\frac{
  \left\langle \frac{\mathrm{d}E^{\mathrm{(in/up)}}}{\mathrm{d}t} \right\rangle_{\mathrm{SN}} 
  -
  \left\langle \frac{\mathrm{d}E^{\mathrm{(in/up)}}}{\mathrm{d}t} \right\rangle_{\mathrm{MST}}
}{
  \left\langle \frac{\mathrm{d}E^{\mathrm{(in/up)}}}{\mathrm{d}t} \right\rangle_{\mathrm{MST}}
}
\right|.
\end{equation}
Fig.~\ref{fig:1} shows that both relative differences remain of order $10^{-7}$ over $5\leq r_{\mathrm{ob}}/M\leq10$, providing a numerical check of the precision required for the calculations below.

\subsection{Orbital evolution}
\label{sec:4.2}

Let $\mathcal{E}(r_{\mathrm{ob}})$ denote the specific orbital energy given in Appendix~\ref{C}, and the corresponding orbital energy is $E_{\mathrm{orb}}=m_s\mathcal{E}$. Adiabatic flux balance gives
\begin{equation}
\label{eq:70}
\frac{\mathrm{d}E_{\mathrm{orb}}(r_{\mathrm{ob}})}{\mathrm{d}t} = -\left( \left\langle \frac{\mathrm{d}E^{\mathrm{in}}}{\mathrm{d}t} \right\rangle 
+ \left\langle \frac{\mathrm{d}E^{\mathrm{up}}}{\mathrm{d}t} \right\rangle \right),
\end{equation}
and hence
\begin{equation}
\label{eq:71}
\frac{\mathrm{d}r_{\mathrm{ob}}}{\mathrm{d}t} = F(r_{\mathrm{ob}}) 
\equiv \left( \frac{\mathrm{d}E_{\mathrm{orb}}(r_{\mathrm{ob}})}{\mathrm{d}t} \right) \!\Big/\! \left( m_s\frac{\mathrm{d}\mathcal{E}(r_{\mathrm{ob}})}{\mathrm{d}r_{\mathrm{ob}}} \right),
\end{equation}
where the derivative of $\mathcal{E}$ is evaluated along the near-equatorial circular-orbit family. The explicit $\mathcal{E}(r)$ is obtained from the construction in Appendix~\ref{C}.

\begin{figure*}[t]
\centering
\includegraphics[width=.43\textwidth]{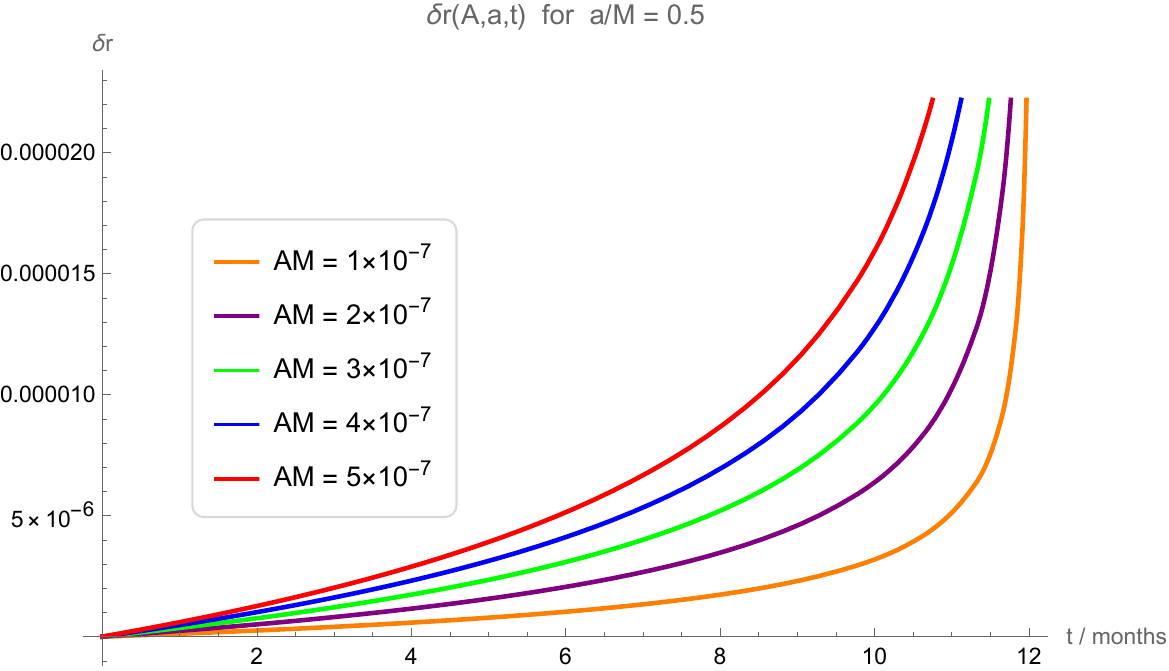}
\qquad
\includegraphics[width=.43\textwidth]{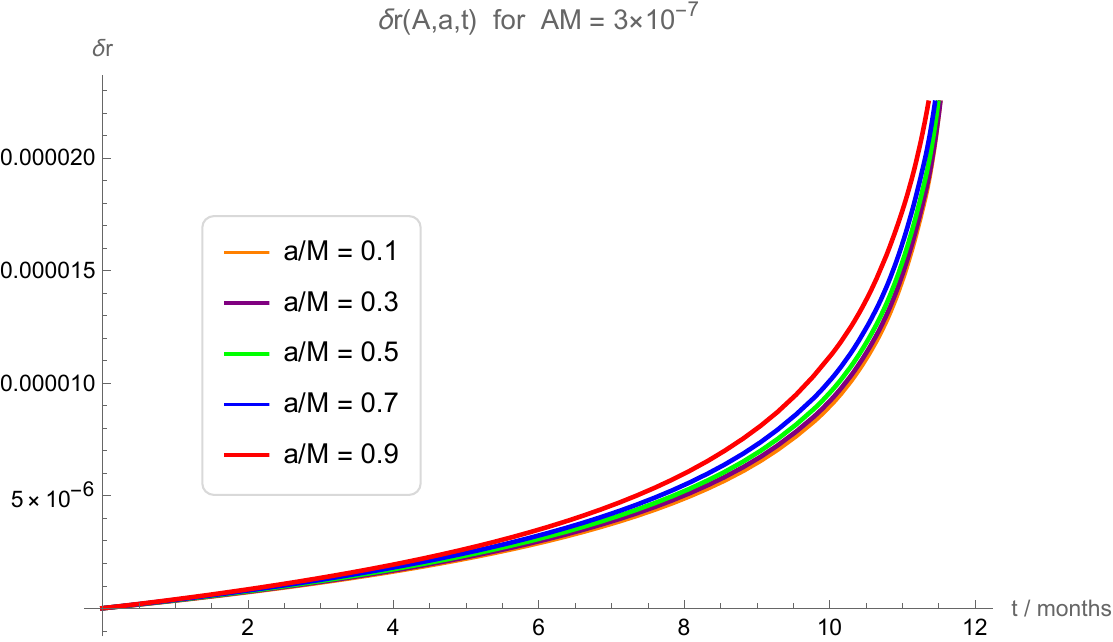}
\caption{\label{fig:2}Acceleration-induced radial difference $\delta r(A,a,t)$, defined in Eq.~\eqref{eq:74}, over a one-year evolution. The left panel varies $A$ at a fixed spin $a/M=0.5$, and the right panel varies $a$ at a fixed acceleration $AM=3 \times 10^{-7}$. We set $s=-2$, $l=2$, the mass ratio $m_s/M=10^{-5}$, and follow the prograde near-equatorial circular orbit.}
\end{figure*}

Based on Eq.~\eqref{eq:71}, the orbital radius and its derivative at any time can be provided by solving the following initial-value problem:
\begin{equation}
\label{eq:72}
\begin{split}
\begin{cases}
\,\dfrac{\mathrm{d}r_{\mathrm{ob}}(t)}{\mathrm{d}t} = F\bigl(r_{\mathrm{ob}}(t)\bigr)
\\[8pt]
\,r_{\mathrm{ob}}(0)=r_0
\end{cases},
\end{split}
\end{equation}
where the orbital radius $r_{\mathrm{ob}}(t)$ is evolved by a forward integration from the initial time $t_0=0\, \mathrm{s}$, and the initial orbital radius $r_0$ is chosen according to Table~\ref{tab:1}. The integration path is
\begin{equation}
\label{eq:73}
t_0  \to T,
\end{equation}
with the total evolution time $T=1\,\mathrm{year}$. The initial-value problem Eq.~\eqref{eq:72} is solved by using Mathematica's \texttt{NDSolve} which controls the step size of numerical integration adaptively. To isolate the acceleration-induced change at fixed spin $a$, we define
\begin{equation}
\label{eq:74}
\delta r(A,a,t) = r_{\mathrm{ob}}(A,a,t) - r_{\mathrm{ob}}(0,a,t).
\end{equation}
The notation $r_{\mathrm{ob}}(A,a,t)$ makes the parameter dependence explicit. In Fig.~\ref{fig:2}, the left panel shows that the magnitude of the acceleration-induced radial displacement grows with $A$ at fixed $a$, whereas the right panel shows that the magnitude increases with $a$ at fixed $A$ by a smaller margin. Because the two panels vary different parameters with different physical units (one unit of $a$ and $A$ corresponds to $4.43 \times 10^{17} \, \mathrm{m}^{2}/\mathrm{s}$ and $6.086 \times 10^{7} \, \mathrm{m}/\mathrm{s}^{2}$ in turn), they should not be interpreted as a direct comparison of the intrinsic strengths of acceleration and spin.

\subsection{Dephasing}
\label{sec:4.3}
The dominant gravitational-wave frequency along the evolving near-equatorial circular orbit is
\begin{equation}
\label{eq:75}
\omega(A,a,r_{\mathrm{ob}}) = 2P(\pi) \Omega_\varphi,
\end{equation}
where the factor $2$ corresponds to the dominant mode $m_0=2$ \cite{Piovano:2020zin,Estelles:2020osj}, and $\Omega_\varphi$ is given in Appendix~\ref{C}. Along the inspiral, we rewrite Eq.~\eqref{eq:75} in a time-dependent form:
\begin{equation}
\label{eq:76}
\omega(A,a,t) = \omega\big( A, a, r_{\mathrm{ob}}(A,a,t) \big).
\end{equation}
We compare the accelerated and Kerr evolutions at the same spin through
\begin{equation}
\label{eq:77}
\delta\omega(A,a,t) = \omega(A,a,t) - \omega(0,a,t).
\end{equation}
The accumulated phase and the dephasing between the two evolutions are then
\begin{equation}
\label{eq:78}
\begin{split}
\Phi(A,a,t)&=\int_0^t\omega(A,a,t')\,dt',\\
\delta\Phi(A,a,t)&=\left|\int_0^t\delta\omega(A,a,t')\,dt'\right|.
\end{split}
\end{equation}
Throughout the evolution, the dephasing accumulated over a fixed time interval is most sensitive to the late inspiral near the innermost stable circular orbit (ISCO) \cite{Finn:2000gwf}. We therefore adjust the initial radius $r_0$ for each spin $a$ according to Table~\ref{tab:1}, so that the orbit approaches but does not reach the ISCO after the observation time $T=1\,\mathrm{year}$.

For the Kerr case, Table~\ref{tab:1} specifies the initial orbital radius $r_0$ for different $a$ and the corresponding distance $\Delta r_{\rm K}$ between the final evolved position after one year and the ISCO. The distance $\Delta r_{\rm K}$ is defined as
\begin{equation}
\label{stability-buffer-Kerr}
\Delta r_{\rm K}(a)
\equiv r_{\rm ob}(0,a,T)-r_{\rm ISCO}(0,a),
\end{equation}
where $r_{\rm ISCO}(A,a)$ is the ISCO radius, and the minimum distance $\Delta r_{\rm K}(a)$ in  Table~\ref{tab:1} is $0.021$. For the acceleration-corrected case, the distance is defined as
\begin{equation}
\label{stability-buffer-Cm-1}
\Delta r_{\rm C}(A,a)
\equiv r_{\rm ob}(A,a,T)-r_{\rm ISCO}(A,a),
\end{equation}
and it can be rewritten by $\Delta r_{\rm K}(a)$ as
\begin{equation}
\label{stability-buffer-Cm-2}
\begin{split}
\Delta r_{\rm C}(A,a)=&\;\Delta r_{\rm K}(a) + \delta r(A,a,T)\\
&- A^2 r_2(a) + \mathcal{O}(A^3),
\end{split}
\end{equation}
 where $r_{\rm ISCO}(A,a)-r_{\rm ISCO}(0,a)=A^2 r_2(a) + \mathcal{O}(A^3)$, and $r_2(a)$ is given in Appendix~\ref{C}. From the above equation, we can see that the acceleration-corrected  distance needs to be adjusted by two displacements induced by the acceleration. One is the evolution displacement $\delta r(A,a,T)$. In Fig.~\ref{fig:2}, it can be seen that the final evolved positions involving the acceleration lie farther out than their Kerr counterparts, and it means $\delta r(A,a,T)>0$. The other one is the ISCO displacement $r_{\rm ISCO}(A,a)-r_{\rm ISCO}(0,a)$. In Appendix~\ref{C}, our second-order geodesic analysis shows that the acceleration correction leads to an outward ISCO displacement in the range $a/M>0.075$, and the displacement is $A^2 r_2(a) + \mathcal{O}(A^3)$. Therefore, for the acceleration-corrected case with the parameters adopted in our study, the distance $\Delta r_{\rm C}(A,a)$ exceeds $0.021- A^2 r_2(a) + \mathcal{O}(A^3)$, which is clearly a positive value. This means that using the initial radius in Table~\ref{tab:1} can ensure that no waveform is evolved exactly to the acceleration-corrected ISCO and the evolution is within a safe buffer zone. Besides, the results near the endpoint should still be interpreted as an adiabatic circular-orbit estimate, rather than an accurate model of the transition and plunge.

\begin{table*}[t]\small
\centering
\setlength{\tabcolsep}{20pt}
\renewcommand\arraystretch{2}
\begin{tabular}{|c|c|c|c|c|c|c|}
\hline
{$a/M$} & $0.1$ & $0.3$ & $0.5$ & $0.7$ & $0.9$\\
\hline
{$r_0/M$} & $8.96$ & $8.45$ & $7.92$ & $7.36$ & $6.73$\\
\hline
{$\Delta r_{\rm K}/M$} & $0.021$ & $0.217$ & $0.251$ & $0.267$ & $0.288$\\
\hline
\end{tabular}
\caption{\label{tab:1} Initial orbital radius $r_0$ for different $a$ and the corresponding distance $\Delta r_{\rm K}$ between the final evolved position after one year and the ISCO. We set $AM=0$, $s=-2$, $l=2$, the mass ratio $m_s/M=10^{-5}$, and follow the prograde equatorial circular orbit.}
\end{table*}

\begin{figure*}[t]
\centering
\includegraphics[width=.43\textwidth]{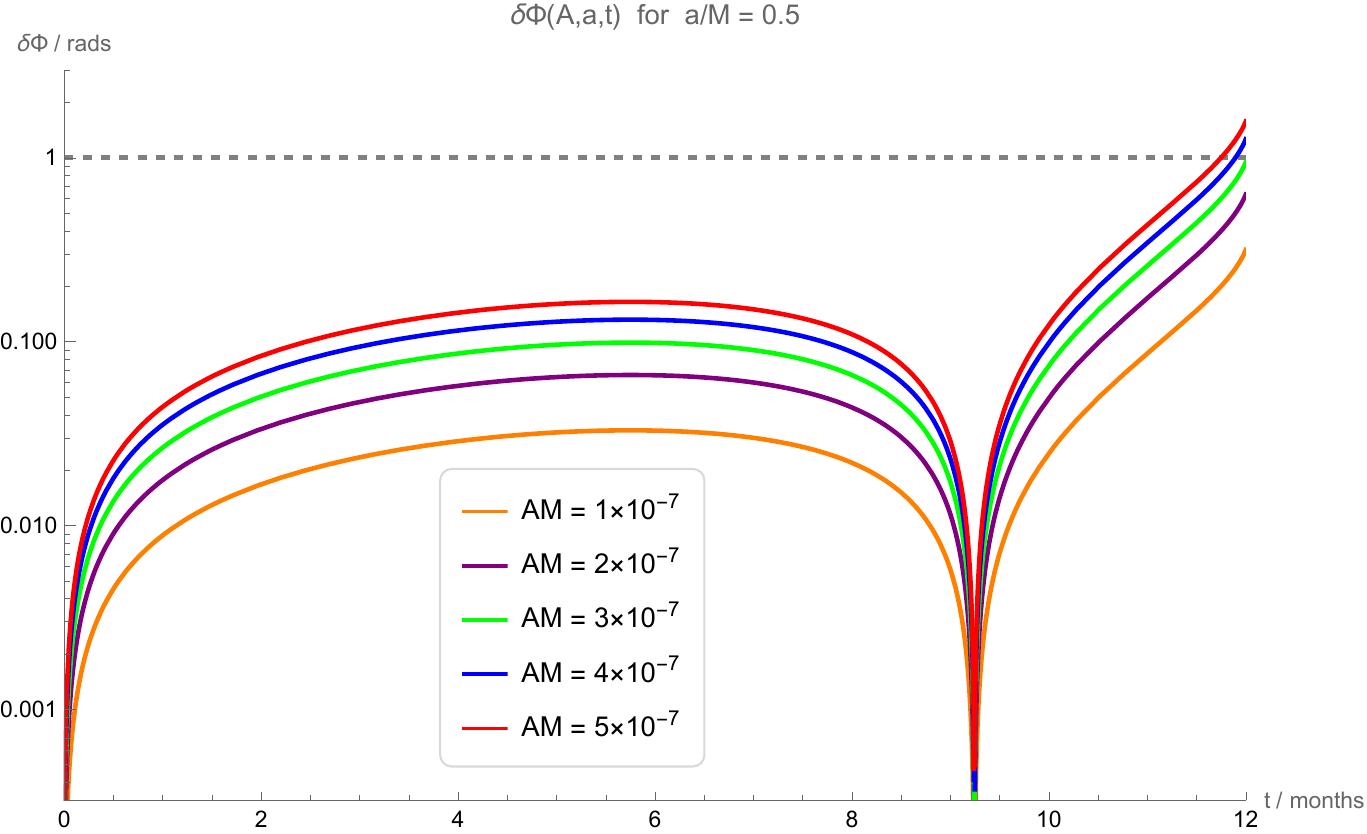}
\qquad
\includegraphics[width=.43\textwidth]{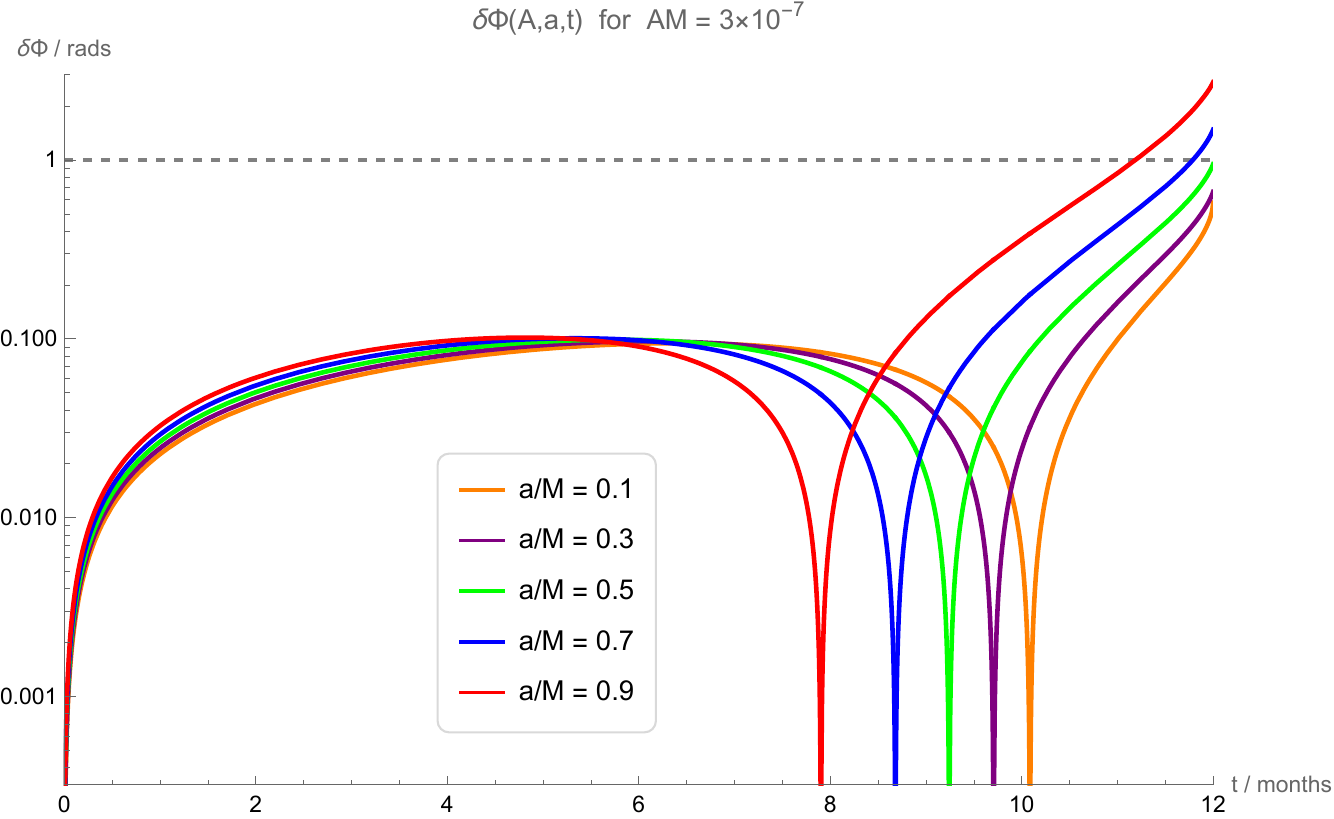}
\caption{\label{fig:3}Accumulated dephasing $\delta\Phi(A,a,t)$, defined in Eq.~\eqref{eq:78}, over one year. The left panel varies $A$, the right panel varies $a$, and the gray horizontal line marks the heuristic $1\,\mathrm{rad}$ benchmark. We set $s=-2$, $l=2$, the mass ratio $m_s/M=10^{-5}$, and follow the prograde near-equatorial circular orbit.}
\end{figure*}

Fig.~\ref{fig:3} shows the accumulated dephasing for several accelerations and spins. The absolute dephasing is nonmonotonic: it initially grows, decreases to a valley, and then grows again. At fixed spin, the late-time magnitude of the dephasing increases with $A$, while increasing $a$ shifts the valley to an earlier time. We use $1\,\mathrm{rad}$ as a heuristic phase-accuracy benchmark \cite{Datta:2020tha}, rather than as a detector-specific distinguishability criterion. In Fig.~\ref{fig:3}, for $a/M=0.5$, the $AM=3\times10^{-7}$ curve remains below this benchmark after one year, whereas the $AM=4\times10^{-7}$ curve exceeds it. For $AM=3\times10^{-7}$, increasing the spin to $a/M=0.7$ also brings the dephasing to above $1\,\mathrm{rad}$. It means that the accumulated dephasing becomes increasingly significant for larger acceleration and spin. These results quantify the secular phase imprint within the approximations of the present waveform model.

\begin{figure*}[t]
\centering
\includegraphics[width=.43\textwidth]{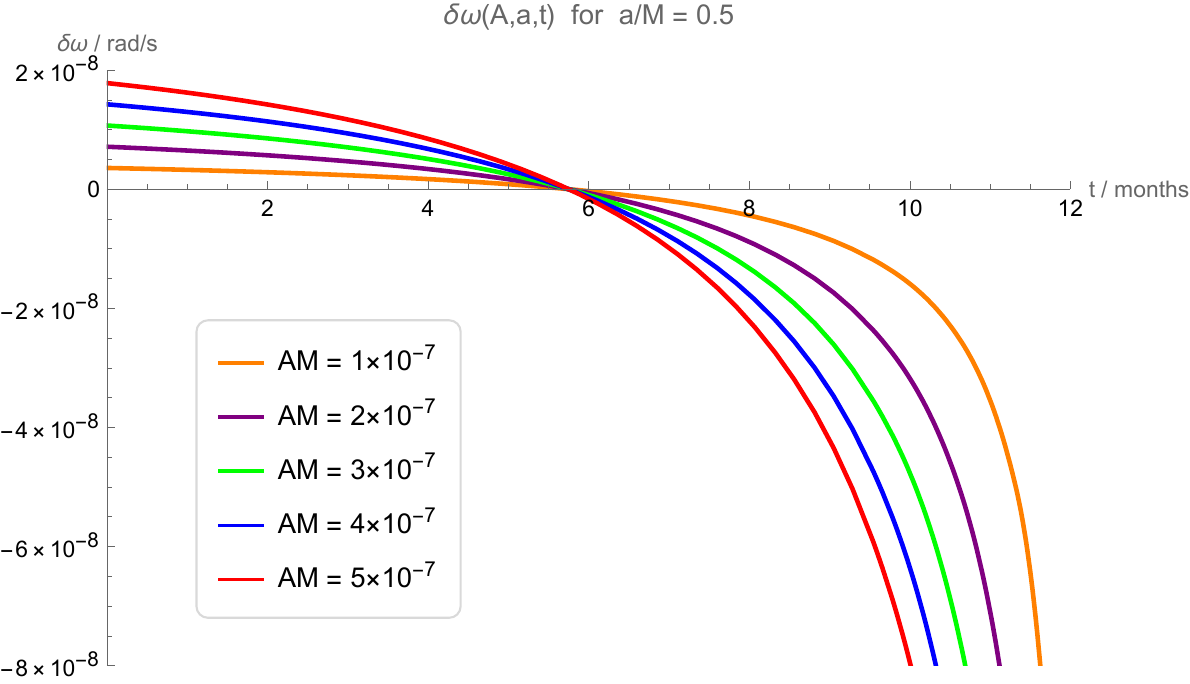}
\qquad
\includegraphics[width=.43\textwidth]{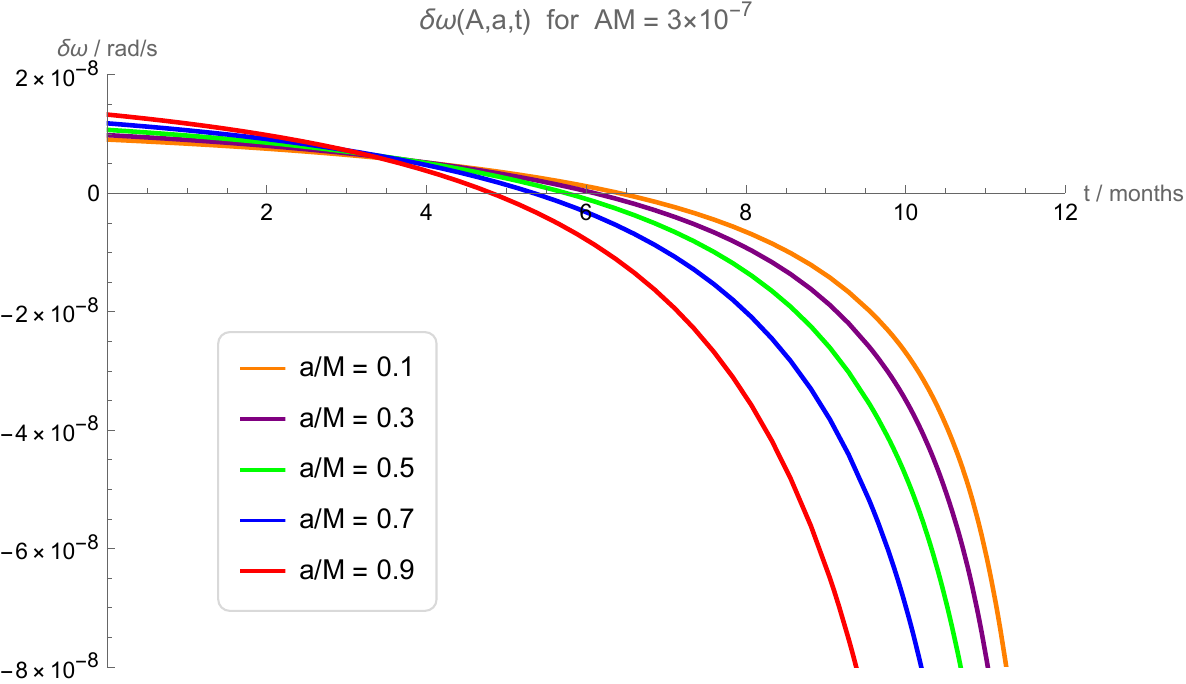}
\caption{\label{fig:4}Frequency difference $\delta\omega(A,a,t)$, defined in Eq.~\eqref{eq:77}, over one year. The left panel varies $A$, and the right panel varies $a$. The sign change produces the valley in the absolute accumulated dephasing shown in Fig.~\ref{fig:3}. We set $s=-2$, $l=2$, the mass ratio $m_s/M=10^{-5}$, and follow the prograde near-equatorial circular orbit. }
\end{figure*}

The origin of this nonmonotonic behavior in Fig.~\ref{fig:3} is visible in Fig.~\ref{fig:4}. The frequency difference $\delta\omega$ changes sign during the inspiral, so positive and negative contributions partially cancel in the time integral in Eq.~\eqref{eq:78}. The valley occurs when the signed accumulated integral passes through zero. Subsequently, the absolute value of the integral then begins to grow again. The zero crossing of $\delta\omega$ depends only weakly on $A$ over the plotted range but occurs earlier as $a$ increases, explaining the corresponding motion of the valley in Fig.~\ref{fig:3}.

\begin{figure*}[t]
\centering
\includegraphics[width=.43\textwidth]{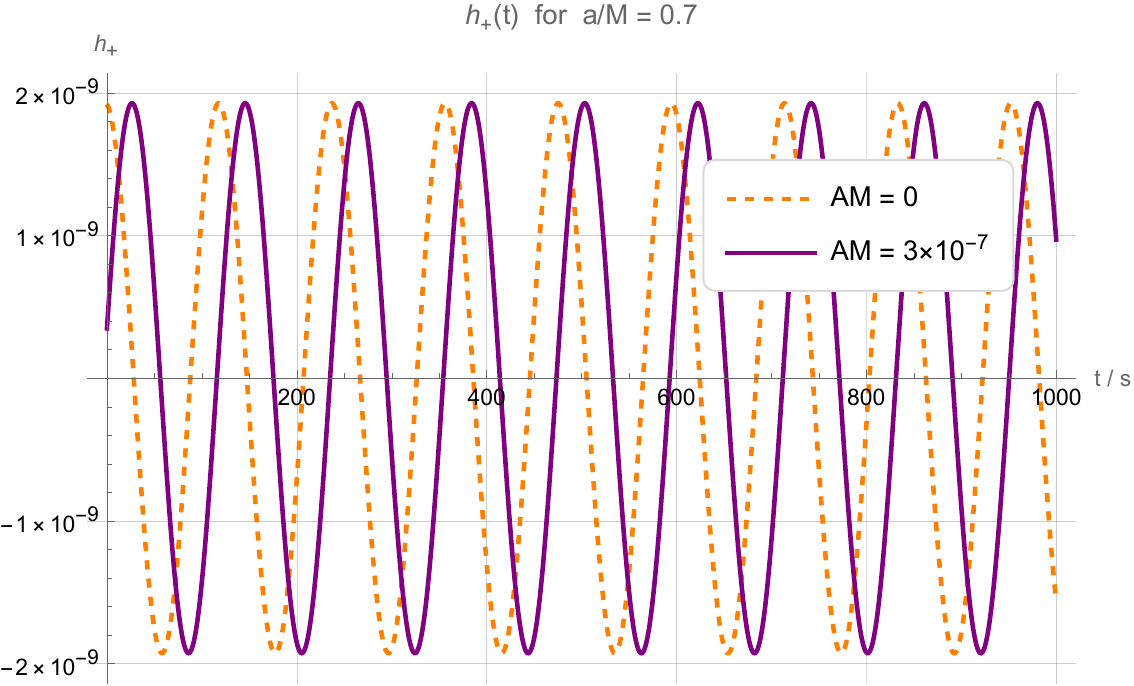}
\qquad
\includegraphics[width=.43\textwidth]{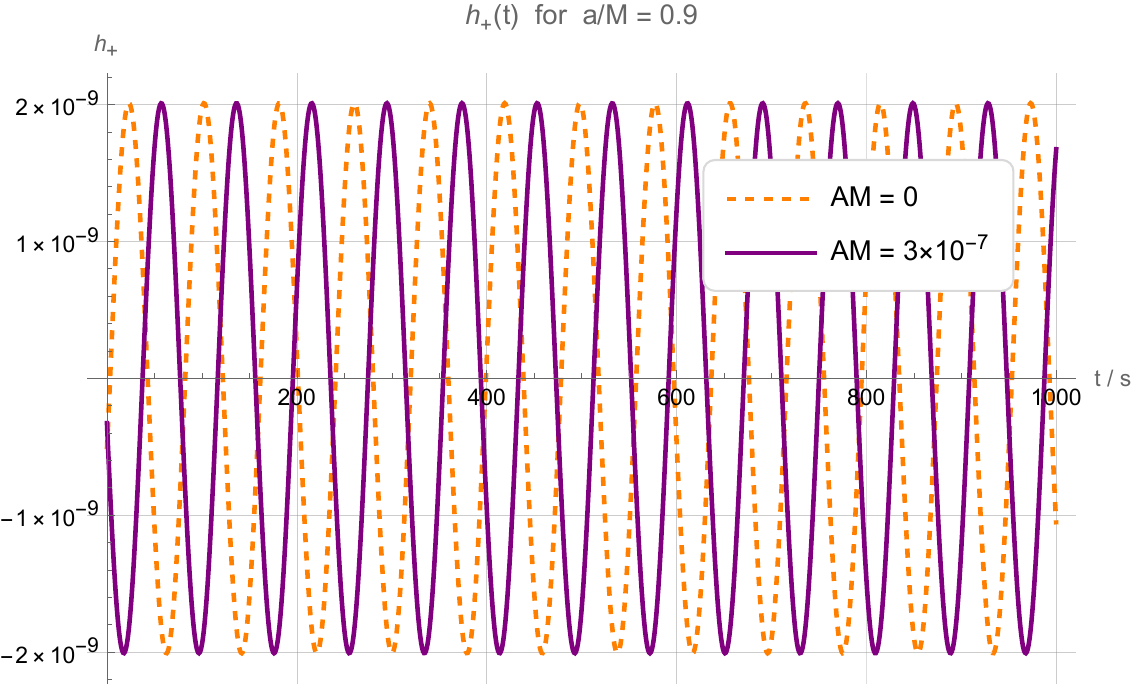}
\caption{\label{fig:5}Plus-polarization waveform $h_+(t)$ from Eq.~\eqref{eq:79} during the last $1000\,\mathrm{s}$ of a one-year evolution, for $a/M=0.7$ (left) and $a/M=0.9$ (right). We set $\theta_{\mathrm{obs}}=\pi/2$, $\varphi_{\mathrm{obs}}=0$, $r_\infty=10^3 M\ll r_A$, $s=-2$, $l=2$, the mass ratio $m_s/M=10^{-5}$, and follow the prograde near-equatorial circular orbit.}
\end{figure*}

For a direct illustration of the phase offset, Fig.~\ref{fig:5} shows the plus-polarization waveform during the last $1000\,\mathrm{s}$ of the one-year evolution. In the intermediate Kerr-like wave zone, the leading waveform is written as \cite{Mino:1997bx,Hughes:1999bq,Sasaki:2003xr}
\begin{equation}
\label{eq:79}
\begin{split}
h_+(t)-ih_\times(t)&=
-\frac{2Z^{\mathrm{in}}S(\theta_{\mathrm{obs}}) e^{-i\Phi(A,a,t)+im\varphi_{\mathrm{obs}}}}
{r_\infty\sqrt{2\pi}\,\omega(A,a,t)^2},
\end{split}
\end{equation}
where $(\theta_{\mathrm{obs}},\varphi_{\mathrm{obs}})$ are the fixed angular coordinates of the observer. They must be distinguished from the time-dependent particle coordinate $(\theta_{\mathrm{ob}}(t),\varphi_{\mathrm{ob}}(t) )$ used in the radial source term Eq.~\eqref{eq:23}. We set $\theta_{\mathrm{obs}}=\pi/2$ and $\varphi_{\mathrm{obs}}=0$ to evaluate the waveform at $r_\infty/M=1000$. In Fig.~\ref{fig:5}, the orange dashed and purple solid curves correspond to $AM=0$ and $AM=3\times10^{-7}$, respectively. The late-time phase offset is approximately one quarter of a cycle for $a/M=0.7$ and one half of a cycle for $a/M=0.9$. The higher-spin waveform also oscillates more rapidly because the late inspiral reaches smaller radius and hence attains higher orbital frequency.


\clearpage
\section{Conclusions and discussion}
\label{sec:5}

We have developed a small-acceleration black-hole perturbation framework for EMRIs in the spinning C-metric spacetime. Because this background is vacuum and Petrov type D away from its conical axis, its spin-$-2$ perturbation equation admits the Teukolsky formulation and separates into radial and angular sectors. We derived the source term of the radial Teukolsky equation, constructed a normalized small-$A$ angular solution that remains well scaled in the Kerr limit, and calculated the homogeneous radial solutions using the Sasaki--Nakamura transformation. Combining these ingredients with a Green-function construction yields the waveform amplitudes at the event horizon and at the intermediate Kerr-like extraction surface. Each part of the formulation reduces continuously to the corresponding Kerr result as $A\to0$.

The absence of north--south reflection symmetry means that the relevant circular orbit is displaced from $\theta=\pi/2$. We determined a near-equatorial circular orbit by examining its coupled radial--polar stability. The determinant and trace of the Hessian receive no linear correction at the Kerr ISCO, so the shift of the marginally stable orbit first appears at $\mathcal O(A^2)$. Our second-order geodesic calculation shows that there is an outward ISCO displacement induced by the acceleration in the spin range used for the inspiral calculations. For numerical stability, our numerical evolutions terminate within a safe buffer zone rather than at the shifted ISCO.

Using the dominant $l=2$ radiation, we find that the secular phase imprint grows with both the acceleration and the prograde spin. The accumulated dephasings show a characteristic nonmonotonic trend, which is caused by a sign change in the acceleration-induced frequency difference. For $AM=3\times10^{-7}$, $a/M=0.7$, $r_\infty/M=10^3$, $M=10^6M_\odot$, and $m_s/M=10^{-5}$, the accumulated dephasing over one year slightly exceeds $1$ rad. This provides a quantitative indication that even a weak acceleration can become relevant through the long coherent phase evolution of an EMRI.

The present results should be interpreted within the controlled approximation scheme adopted here. The spinning C metric is not globally asymptotically flat, so the flux is extracted instead in an intermediate region satisfying $M/r_\infty\ll1$ and $Ar_\infty\ll1$, and all radiative quantities are truncated at first order in $A$. We have also retained only the dominant $l=2$ modes and used the Kerr-form horizon-flux factor, while the transition from adiabatic inspiral to plunge is outside the present model. A more complete treatment should impose the physical boundary condition at the acceleration horizon, include higher multipoles and acceleration corrections to the horizon flux, and connect the circular inspiral to a transition-to-plunge evolution. Within the regime established here, however, the spinning C metric provides a tractable non-Kerr laboratory in which the acceleration of the central source produces a calculable order-radian secular imprint on EMRI gravitational waves.

\begin{acknowledgments}
This work made use of the Black Hole Perturbation Toolkit. The work was supported in part by the National Natural Science Foundation of China under Grant No.~12205104, No. 12347140 and No. 12405059. T.Zi is also supported by the Startup Fund for Advanced Talents of Nanchang University with Grant No. 28170256.
\end{acknowledgments}

\appendix

\section{Newman--Penrose quantities}
\label{A}
For the spinning C metric, we use the following Newman--Penrose tetrad:
\begin{equation}
\label{eq:A1}
\begin{split}
l^{\mu} &= \Omega^2 \left( \frac{r^2 + a^2}{Q},  1,  0,  \frac{a}{Q} \right),
\\
n^{\mu} &= \frac{1}{2} \Sigma \left( r^2 + a^2, -Q, 0, a \right),
\\
m^{\mu} &= \frac{\Omega}{\sqrt{2} P (r \!+\! i a \cos\theta)} \left( i a \sin\theta, 0, P, \frac{i} {\sin\theta} \right),
\\
\bar{m}^{\mu} &= \frac{\Omega}{\sqrt{2} P (r \!-\! i a \cos\theta)}  \left(\!-\!ia\sin\theta, 0, P, \frac{\!-\!i} {\sin\theta} \right),
\end{split}
\end{equation}
with the normalization:
\begin{equation}
\label{eq:A2}
l^{\mu} n_{\mu} = -m^{\mu} \bar{m}_{\mu} = -1.
\end{equation}
The 12 Newman--Penrose spin coefficients are defined by
\begin{equation}
\label{eq:A3}
\begin{split}
\kappa &= -m^{\mu} l^{\nu} \nabla_{\nu} l_{\mu}, \quad\;\;\quad
\lambda = -n^{\mu} \bar{m}^{\nu} \nabla_{\nu} \bar{m}_{\mu}, 
\\
\sigma &= -m^{\mu} m^{\nu} \nabla_{\nu} l_{\mu}, \quad\quad
\nu = -n^{\mu} n^{\nu} \nabla_{\nu} \bar{m}_{\mu},
\\
\rho &= -m^{\mu} \bar{m}^{\nu} \nabla_{\nu} l_{\mu}, \quad\quad
\mu = -n^{\mu} m^{\nu} \nabla_{\nu} \bar{m}_{\mu}, 
\\
\tau &= -m^{\mu} n^{\nu} \nabla_{\nu} l_{\mu}, \quad\quad
\varpi = -n^{\mu} l^{\nu} \nabla_{\nu} \bar{m}_{\mu},
\\
\epsilon &= -\frac{1}{2} \left( n^{\mu} l^{\nu} \nabla_{\nu} l_{\mu} + m^{\mu} l^{\nu} \nabla_{\nu} \bar{m}_{\mu} \right),
\\
\gamma &= -\frac{1}{2} \left( n^{\mu} n^{\nu} \nabla_{\nu} l_{\mu} + m^{\mu} n^{\nu} \nabla_{\nu} \bar{m}_{\mu} \right),
\\
\alpha &= -\frac{1}{2} \left( n^{\mu} \bar{m}^{\nu} \nabla_{\nu} l_{\mu} + m^{\mu} \bar{m}^{\nu} \nabla_{\nu} \bar{m}_{\mu} \right),
\\
\beta &= -\frac{1}{2} \left( n^{\mu} m^{\nu} \nabla_{\nu} l_{\mu} + m^{\mu} m^{\nu} \nabla_{\nu} \bar{m}_{\mu} \right),
\end{split}
\end{equation}
and the 5 Weyl scalars by
 \begin{align}
\label{eq:A4}
\psi_0 &= C_{\alpha\beta\mu\nu} l^\alpha m^\beta l^\mu m^\nu, 
\\
\psi_1 &= C_{\alpha\beta\mu\nu} l^\alpha n^\beta l^\mu m^\nu, 
\\
\psi_2 &= C_{\alpha\beta\mu\nu} l^\alpha m^\beta \bar{m}^\mu n^\nu, 
\\
\psi_3 &= C_{\alpha\beta\mu\nu} l^\alpha n^\beta \bar{m}^\mu n^\nu, 
\\
\psi_4 &= C_{\alpha\beta\mu\nu} n^\alpha \bar{m}^\beta n^\mu \bar{m}^\nu,
\end{align}
where the Weyl tensor is
\begin{equation}
\label{eq:A5}
\begin{split}
C_{\alpha\beta\mu\nu}={}&R_{\alpha\beta\mu\nu}
-\frac{2}{n-2}\left(g_{\alpha[\mu}R_{\nu]\beta}
-g_{\beta[\mu}R_{\nu]\alpha}\right)\\
&+\frac{2R}{(n-1)(n-2)}g_{\alpha[\mu}g_{\nu]\beta}.
\end{split}
\end{equation}
Here $n$ is the spacetime dimension, and $n=4$ throughout this work. For completeness, our curvature and antisymmetrization conventions are
\begin{equation}
\label{eq:A6}
\begin{split}
g_{\alpha[\mu} R_{\nu]\beta} &= \frac{1}{2} \left( g_{\alpha\mu} R_{\nu\beta} - g_{\alpha\nu} R_{\mu\beta} \right), 
\\
R_{\beta\mu\nu}^{\alpha} &= \partial_{\mu} \Gamma_{\beta\nu}^{\alpha} - \partial_{\nu} \Gamma_{\beta\mu}^{\alpha} + \Gamma_{\eta\mu}^{\alpha} \Gamma_{\beta\nu}^{\eta} - \Gamma_{\eta\nu}^{\alpha} \Gamma_{\beta\mu}^{\eta}, 
\\
\Gamma_{\beta\mu}^{\alpha} &= \frac{1}{2} g^{\alpha\eta} \left( \partial_{\beta} g_{\mu\eta} + \partial_{\mu} g_{\beta\eta} - \partial_{\eta} g_{\beta\mu} \right).
\end{split}
\end{equation}

\section{Angular eigenvalue and solution}
\label{B}
The separation eigenvalue $\lambda(l,m)$ appearing in the radial and angular equations, Eqs.~\eqref{eq:11} and \eqref{eq:12}, is obtained by solving the following continued fraction equation:
\begin{equation}
\label{eq:B1}
c_2(0) - \frac{c_1(0) c_3(1)}{c_2(1) - \frac{c_1(1) c_3(2)}{c_2(2) - \frac{c_1(2) c_3(3)}{c_2(3) - \cdots}}} = 0,
\end{equation}
where the recurrence coefficients given in Ref.~\cite{Chen:2024rov} are
\begin{equation}
\label{eq:B2}
\begin{split}
c_1(n) =&\; z_+ (1+n)(1+\gamma_a), \\
c_2(n) =&\;  - [z_+ (\gamma_a + \delta_a - 1) + \gamma_a + \epsilon_a - 1] n 
\\
&\;-(z_+ + 1) n^2 - q_a,
\\
c_3(n) =&\; n^2 + (\gamma_a + \delta_a + \epsilon_a - 3)(n-1) 
\\
&\; + \alpha_a \beta_a - 1,
\end{split}
\end{equation}
with
\begin{equation}
\label{eq:B3}
\begin{split}
\gamma_a =&\; 2A_1 + 1,\\
\delta_a =&\; 2A_2 + 1,\\
\epsilon_a =&\; 2A_3 + 1,\\
\alpha_a =&\; 1 + A_1 + A_2 - \frac{\tilde{\omega}}{\sqrt{\tilde{A}^2 - \tilde{a}^2}} \\
&\;+ \frac{m(\tilde{a}^4 + \tilde{a}^2 - 2\tilde{A}^2)}{\sqrt{\tilde{A}^2 - \tilde{a}^2}  P(0) P(\pi)},\\
\beta_a =&\; 1 + s + A_1 + A_2 - \frac{2 A m M}{P(0) P(\pi)},\\
q_a =&\; z_+ (2A_1 A_2 + A_1 + A_2) + 2A_1 A_3 + A_1 + A_3\\
&\;+ A_1^2 + \frac{1}{2}
- \frac{1}{2\tilde{a}^2 (u_+ - u_-)} \Bigg[ -\lambda  + \frac{m^2}{P(\pi)}\\
&\;- (1 - \tilde{a}^2 - m)s - \frac{1}{2}(1 - \tilde{a}^2)\left(s + \frac{m}{P(\pi)}\right)^2\\
&\;- 2\left(s + \frac{m}{P(\pi)}\right)\tilde{\omega} \Bigg],
\end{split}
\end{equation}
where
\begin{equation}
\label{eq:B4}
\begin{split}
A_1 =&\; \frac{1}{2} \left| \frac{m}{P(\pi)} - s \right|, \\
A_2 =&\; \frac{1}{2} \left| \frac{m}{P(\pi)} + s \right|, \\
A_3 =&\; \frac{1}{2} \Bigg\{ s + \frac{m\left[ \tilde{a}^4 + \tilde{a}^2 - 2\tilde{A}\left(\tilde{A} + \sqrt{\tilde{A}^2 - \tilde{a}^2}\right) \right]}{\sqrt{\tilde{A}^2 - \tilde{a}^2}  P(0) P(\pi)}\\
&\;- \frac{\tilde{\omega}}{\sqrt{\tilde{A}^2 - \tilde{a}^2}} \Bigg\}, \\
u_{\pm} =&\; \frac{1}{\tilde{A} \pm \sqrt{\tilde{A}^2 - \tilde{a}^2}},
\end{split}
\end{equation}
and
\begin{equation}
\label{eq:B5}
\tilde{a} = A a, \quad \tilde{A} = A M, \quad \tilde{\omega} = a \omega.
\end{equation}
Here $s=-2$ describes gravitational perturbations. Relative to Ref.~\cite{Chen:2024rov}, the sign multiplying $\lambda$ in $q_a$ is reversed to match the convention adopted in Eqs.~\eqref{eq:13} and \eqref{eq:14}. Because Eq.~\eqref{eq:B1} has multiple roots, we identify the desired branch by continuation from the Kerr eigenvalue $\lambda_0(l,m_0)$ as $A$ is turned on.

Reference~\cite{Chen:2024rov} gives an unnormalized Frobenius-series solution of Eq.~\eqref{eq:12}:
\begin{equation}
\label{eq:B6}
S(z)=z^{A_1}(z-1)^{A_2}(z-z_+)^{A_3}(z-z_\infty)\sum_{n=0}^{\infty} a_n z^n,
\end{equation}
where
\begin{equation}
\label{eq:B7}
\begin{split}
& z_\infty= \frac{1-u_-}{2}, \qquad z_+= \frac{(u_++1)(1-u_-)}{2(u_+-u_-)},\\
& z= \frac{(\cos\theta+1)(1-u_-)}{2(\cos\theta-u_-)}.\\
\end{split}
\end{equation}
The expansion coefficients $\{a_n\}$ satisfy the following three-term recurrence relation:
\begin{equation}
\label{eq:B8}
\begin{split}
&c_1(0) a_1 + c_2(0) a_0 = 0,\\
&c_1(n) a_{n+1} + c_2(n) a_n + c_3(n) a_{n-1} = 0,
\end{split}
\end{equation}
with $c_1(n)$, $c_2(n)$, and $c_3(n)$ from Eq.~\eqref{eq:B2} and $n = 1, 2, 3, \cdots$. The overall scale is arbitrary at this stage, and we choose $a_0=1$.

For $A\ll1$, the factors $(z-z_+)^{A_3}$ and $(z-z_\infty)$ in Eq.~\eqref{eq:B6} generate a severe overall scale hierarchy and lead to a loss of numerical precision during normalization. Because the overall scale of $S$ is arbitrary, we remove this hierarchy by dividing each factor by its limiting value at $A=0$ and $\theta=\pi/2$ to obtain
\begin{equation}
\label{eq:B9}
\begin{split}
&\left(z - z_{+}\right)^{A_3} \to e^{a \omega \cos\theta} \!+\! e^{a \omega \cos\theta} \cos\theta \Big(a M \omega \cos\theta 
\\ &\qquad\qquad\qquad-\! s \sqrt{M^2 \!-\! a^2}\Big) A \!+\! \mathcal{O}(A^2), \\
&\left(z - z_{\infty}\right) \to 1 \!+\! \left(M \!-\! \sqrt{M^2 \!-\! a^2}\right) \cos\theta A \!+\! \mathcal{O}(A^2).
\end{split}
\end{equation}
Besides, to avoid the occurrence of imaginary numbers, we also choose the real branch convention on $0\leq z\leq1$ by replacing
\begin{equation}
\label{eq:B10}
(z-1)^{A_2} \rightarrow (1-z)^{A_2}.
\end{equation}
Applying the rescaled results in Eq.~\eqref{eq:B9} and the branch convention in Eq.~\eqref{eq:B10} to Eq.~\eqref{eq:B6} yields the well-scaled unnormalized solution:
\begin{equation}
\label{eq:B11}
S_0(z) = z^{A_1} (1-z)^{A_2} Z_1 Z_2 \sum_{n=0}^{\infty} a_n z^n,
\end{equation}
where $Z_1$ and $Z_2$ have the first-order expansions
\begin{equation}
\label{eq:B12}
\begin{split}
Z_1 =&\; e^{a\omega \cos\theta} + e^{a\omega \cos\theta} \cos\theta \Big( a M \omega \cos\theta \\
&\;- s \sqrt{M^2 - a^2} \Big) A, \\
Z_2 =&\; 1 + \left( M - \sqrt{M^2 - a^2} \right) \cos\theta A.
\end{split}
\end{equation}
The normalized angular solution is therefore
\begin{equation}
\label{eq:B13}
S(l,m,z) = \frac{1}{\sqrt{\int_{-1}^{1} \left[ S_{0}(z) \right]^2  \mathrm{d}(\cos\theta)}} S_{0}(z),
\end{equation}
where $(l,m)$ label the eigenvalue branch $\lambda(l,m)$ selected above.

\twocolumngrid
\section{Near-equatorial circular orbit}
\label{C}
In this appendix, we construct a near-equatorial circular orbit by examining its coupled radial--polar stability described by an effective potential. We start from the timelike geodesic normalization condition for the spinning C metric:
\begin{equation}
\label{eq:C1}
-1=g_{rr}\dot r^2+g_{tt}\dot t^2+g_{\theta\theta}\dot\theta^2
+2g_{t\varphi}\dot t\dot\varphi+g_{\varphi\varphi}\dot\varphi^2,
\end{equation}
where the dot denotes differentiation with respect to proper time. Stationarity and axisymmetry of the spinning C metric provide the conserved specific energy $\mathcal{E}$ and specific angular momentum $\mathcal{L}$:
\begin{equation}
\label{eq:C2}
\mathcal{E}=-g_{tt}\dot t-g_{t\varphi}\dot\varphi,
\qquad
\mathcal{L}=g_{t\varphi}\dot t+g_{\varphi\varphi}\dot\varphi,
\end{equation}
leading to
\begin{equation}
\label{eq:C3}
\dot t=\frac{\mathcal{E}g_{\varphi\varphi}+\mathcal{L}g_{t\varphi}}
{g_{t\varphi}^2-g_{tt}g_{\varphi\varphi}},
\qquad
\dot\varphi=-\frac{\mathcal{E}g_{t\varphi}+\mathcal{L}g_{tt}}
{g_{t\varphi}^2-g_{tt}g_{\varphi\varphi}},
\end{equation}
and hence
\begin{equation}
\label{eq:C4}
\Omega_\varphi\equiv\frac{\dot\varphi}{\dot t}
=-\frac{\mathcal{E}g_{t\varphi}+\mathcal{L}g_{tt}}
{\mathcal{E}g_{\varphi\varphi}+\mathcal{L}g_{t\varphi}}.
\end{equation}

Following \cite{Tahara:2024pot}, we define the two-dimensional effective potential:
\begin{equation}
\label{eq:C5}
\begin{split}
V_{\mathrm{eff}}(r,\theta;\mathcal{E},\mathcal{L})
={}&1+g_{tt}\dot t^2+2g_{t\varphi}\dot t\dot\varphi
+g_{\varphi\varphi}\dot\varphi^2
\\
={}&-g_{rr}\dot r^2-g_{\theta\theta}\dot\theta^2.
\end{split}
\end{equation}
All derivatives of $V_{\mathrm{eff}}$ below are taken at fixed $(\mathcal{E},\mathcal{L})$. A circular orbit at constant $(r_{\mathrm{ob}},\theta_{\mathrm{ob}})$ satisfies
\begin{equation}
\label{eq:C6}
V_{\mathrm{eff}}(r_{\mathrm{ob}},\! \theta_{\mathrm{ob}})
\!=\! \partial_r\! V_{\mathrm{eff}}(r_{\mathrm{ob}} ,\!  \theta_{\mathrm{ob}})
\!=\! \partial_\theta\! V_{\mathrm{eff}}(r_{\mathrm{ob}} ,\! \theta_{\mathrm{ob}})\!=\!0.
\end{equation}
The linearized radial and polar motions are governed by the metric-weighted Hessian of $V_{\mathrm{eff}}$. The circular orbit is stable if the radial and angular eigenvalues of the Hessian are positive. This is equivalent to requiring the determinant $\mathcal{D}_{\mathrm{ob}}$ and trace $\mathcal{T}_{\mathrm{ob}}$ of the Hessian being positive, namely:
\begin{equation}
\label{eq:C7}
\mathcal{D}_{\mathrm{ob}}
\equiv
\frac{\partial_r^2V_{\mathrm{eff}}\,\partial_\theta^2V_{\mathrm{eff}}
-(\partial_r\partial_\theta V_{\mathrm{eff}})^2}
{g_{rr}g_{\theta\theta}}>0,
\end{equation}
\begin{equation}
\label{eq:C8}
\mathcal{T}_{\mathrm{ob}}
\equiv
\frac{\partial_r^2V_{\mathrm{eff}}}{g_{rr}}
+\frac{\partial_\theta^2V_{\mathrm{eff}}}{g_{\theta\theta}}>0.
\end{equation}

Rather than assuming stability, we seek a regular perturbative continuation of a Kerr equatorial circular orbit. For small $A$, let
\begin{equation}
\label{eq:C9}
\begin{split}
\theta_{\mathrm{ob}} =&\; \frac{\pi}{2}+A f_1(r_{\mathrm{ob}})+A^2f_2(r_{\mathrm{ob}})+\mathcal{O}(A^3),\\
\mathcal{E} =&\; \mathcal{E}_{\mathrm{K}} + A\mathcal{E}_1 + A^2\mathcal{E}_2 + \mathcal{O}(A^3),\\
\mathcal{L}=&\; \mathcal{L}_{\mathrm{K}} + A\mathcal{L}_1 + A^2\mathcal{L}_2 + \mathcal{O}(A^3).
\end{split}
\end{equation}
At zeroth order, Eq.~\eqref{eq:C6} gives the Kerr circular-orbit quantities. Introducing $\sigma=+1$ for prograde orbits and $\sigma=-1$ for retrograde orbits, they can be written as
\begin{equation}
\label{eq:C10}
\mathcal{E}_{\mathrm{K}}^{(\sigma)}
=\frac{r_{\mathrm{ob}}^{3/2}-2Mr_{\mathrm{ob}}^{1/2}+\sigma aM^{1/2}}
{r_{\mathrm{ob}}^{3/4}\left(r_{\mathrm{ob}}^{3/2}-3Mr_{\mathrm{ob}}^{1/2}
+2\sigma aM^{1/2}\right)^{1/2}},
\end{equation}
\begin{equation}
\label{eq:C11}
\mathcal{L}_{\mathrm{K}}^{(\sigma)}
=\frac{\sigma M^{1/2}\left(r_{\mathrm{ob}}^2-2\sigma aM^{1/2}r_{\mathrm{ob}}^{1/2}+a^2\right)}
{r_{\mathrm{ob}}^{3/4}\left(r_{\mathrm{ob}}^{3/2}-3Mr_{\mathrm{ob}}^{1/2}
+2\sigma aM^{1/2}\right)^{1/2}}.
\end{equation}
The prograde orbits are used in our following calculations. Expanding all three conditions in Eq.~\eqref{eq:C6} through first order in $A$, the normalization and radial conditions give
$\mathcal{E}_1=\mathcal{L}_1=0$, while the polar condition determines the displacement from the equatorial plane. The polar condition gives
\begin{equation}
\label{eq:C12}
f_1(r_{\mathrm{ob}})=\frac{f_N(r_{\mathrm{ob}})}{f_D(r_{\mathrm{ob}})},
\end{equation}
with
\begin{equation}
\label{eq:C13}
\begin{split}
f_N(r_{\mathrm{ob}})=&\; r_{\mathrm{ob}}^4 - 3M r_{\mathrm{ob}}^3 + 2a M^{1/2} r_{\mathrm{ob}}^{5/2} + M^2 r_{\mathrm{ob}}^2\\
&\;- 2a M^{3/2} r_{\mathrm{ob}}^{3/2} + a^2 M r_{\mathrm{ob}},\\
f_D(r_{\mathrm{ob}})=&\;M r_{\mathrm{ob}}^2 - 4a M^{3/2} r_{\mathrm{ob}}^{1/2} + 3a^2 M.
\end{split}
\end{equation}
Thus, $\theta_{\mathrm{ob}}-\pi/2=\mathcal{O}(A)$, whereas the first corrections to $\mathcal{E}$ and $\mathcal{L}$ occur at $\mathcal{O}(A^2)$. At the next order, the three coefficients $f_2(r_{\mathrm{ob}})$, $\mathcal{E}_2$, and $\mathcal{L}_2$ are obtained by setting the $\mathcal{O}(A^2)$ parts of $V_{\mathrm{eff}}$, $\partial_rV_{\mathrm{eff}}$, and $\partial_\theta V_{\mathrm{eff}}$ to zero. Their explicit rational expressions are lengthy and are not displayed, but they are retained in the evaluation of the stability invariants below.

Using the circular orbit results Eqs.~\eqref{eq:C10}, \eqref{eq:C11} and \eqref{eq:C12}, the two stability invariants in Eqs.~\eqref{eq:C7} and \eqref{eq:C8} are calculated:
\begin{equation}
\label{eq:C14}
\begin{split}
\mathcal{D}_{\mathrm{ob}}(r_{\mathrm{ob}},\!a,\!A)
\!=& \mathcal{D}_{\mathrm K}(r_{\mathrm{ob}},\!a)
\!+\! A^2\mathcal{D}_2(r_{\mathrm{ob}},\!a) \!+\! \mathcal{O}(A^3),\\
\mathcal{T}_{\mathrm{ob}}(r_{\mathrm{ob}},\!a,\!A)
\!=& \frac{4M}{r_{\mathrm{ob}}^3} \!+\! A^2\mathcal{T}_2(r_{\mathrm{ob}},\!a)
\!+\! \mathcal{O}(A^3),
\end{split}
\end{equation}
with
\begin{equation}
\label{eq:C15}
\begin{split}
\mathcal{D}_{\mathrm K}(r_{\mathrm{ob}},\!a)
=& \frac{4M^2\left(3a^2 \!-\! 4a\sqrt{Mr_{\mathrm{ob}}} \!+\! r_{\mathrm{ob}}^2\right)}
{r_{\mathrm{ob}}^7\left(2a\sqrt{M} \!+\! r_{\mathrm{ob}}\sqrt{r_{\mathrm{ob}}} \!-\! 3M\sqrt{r_{\mathrm{ob}}}\right)^2}\\
& \times \left( 8a\sqrt{Mr_{\mathrm{ob}}} \!-\! 3a^2 \!+\! r_{\mathrm{ob}}^2 \!-\! 6Mr_{\mathrm{ob}}\right)\!.
\end{split}
\end{equation}
The terms linear in $A$ vanish in the above two stability invariants. The zero-order terms in Eq.~\eqref{eq:C14} are exactly the Kerr stability invariants which are positive and bounded away from zero, Eq.~\eqref{eq:C14} therefore establishes a stable near-equatorial continuation for sufficiently small $A$. The two-dimensional Hessian criterion itself remains valid at marginal stability, and we therefore can resolve the displacement of the ISCO by retaining the displayed second-order terms.

For the resulting circular orbit, its four-velocity is
\begin{equation}
\label{eq:C16}
u^\mu=(\dot t,0,0,\dot\varphi),
\end{equation}
where $\dot t$ and $\dot\varphi$ are determined by inserting
$\mathcal{E}=\mathcal{E}_{\mathrm{K}}+\mathcal{O}(A^2)$ and
$\mathcal{L}=\mathcal{L}_{\mathrm{K}}+\mathcal{O}(A^2)$ into Eq.~\eqref{eq:C3}. The orbital frequency $\Omega_\varphi$ and the modal frequency $\omega$ are then given by
\begin{equation}
\label{eq:C17}
\begin{split}
\Omega_\varphi&=\frac{\dot\varphi}{\dot t}=\frac{\sqrt{M}}{a\sqrt{M}+r_{\mathrm{ob}}^{3/2}}
+\mathcal{O}(A^2),\\
\omega&=m\Omega_\varphi=m_0P(\pi)\Omega_\varphi.
\end{split}
\end{equation}
The orbital frequency has no correction linear in $A$, while at fixed integer $m_0$ the modal frequency also contains the acceleration dependence carried by the allowed azimuthal number $m$. 

Finally, we study the ISCO displacement induced by the acceleration. Let $r_I=r_{\rm ISCO}(0,a)$ denote the prograde Kerr ISCO. It satisfies $\mathcal{D}_{\mathrm K}(r_I,a)=0$, while $\partial_r\mathcal{D}_{\mathrm K}(r_I,a)>0$ for the nonextremal branch considered here. At the old Kerr ISCO radius, Eq.~\eqref{eq:C14} gives
\begin{equation}
\label{eq:C18}
\begin{split}
\mathcal{D}_{\mathrm{ob}}(r_I,a,A)
&=A^2\mathcal{D}_2(r_I,a)+\mathcal{O}(A^3),\\
\mathcal{T}_{\mathrm{ob}}(r_I,a,A)
&=\frac{4M}{r_I^3}+A^2\mathcal{T}_2(r_I,a)
+\mathcal{O}(A^3),\\
\lambda_r(r_I,a,A)
&=\frac{A^2 r_I^3}{4M}\mathcal{D}_2(r_I,a)
+\mathcal{O}(A^3),
\end{split}
\end{equation}
where $\lambda_r$ is the radial eigenvalue that vanishes in the Kerr limit. Direct evaluation of the exact second-order coefficients over $10^{-3}\leq a/M\leq1$ gives $\mathcal{T}_2(r_I,a)>0$ throughout the prograde branch. Together with $4M/r_I^3>0$, this ensures $\mathcal{T}_{\mathrm{ob}}(r_I,a,A)>0$ at the order considered. Over the same branch, $\mathcal{D}_2(r_I,a)<0$ for $a/M>0.075$. Hence the orbit located at the old Kerr ISCO radius acquires one negative radial eigenvalue in that spin range. The true ISCO is instead
\begin{equation}
\label{eq:C19}
r_{\mathrm{ISCO}}(A,a) =r_I+A^2r_2(a)+\mathcal{O}(A^3),
\end{equation}
with
\begin{equation}
\label{eq:C20}
r_2(a) =-\frac{\mathcal{D}_2(r_I,a)}
{\partial_r\mathcal{D}_{\mathrm K}(r_I,a)}>0
\quad (a/M>0.075).
\end{equation}
Thus, throughout the spin range $0.1\leq a/M\leq0.9$ used in Sec.~\ref{sec:4.3}, acceleration shifts the ISCO outward at second order. At the shifted radius, $\mathcal{D}_{\mathrm{ob}}=0$ and $\mathcal{T}_{\mathrm{ob}}>0$, as required for a marginally stable circular orbit. For numerical stability, our numerical evolutions terminate within a safe buffer zone rather than at the shifted ISCO. This targeted $\mathcal{O}(A^2)$ geodesic calculation is used only to identify the stability boundary, and it does not change the first-order truncation adopted for the perturbation amplitudes, fluxes, and inspiral waveforms.

\section{Source-term coefficients}
\label{D}
The coefficients $A_{nn0}$, $A_{\bar mn0}$, $A_{\bar m\bar m0}$, $A_{\bar mn1}$, $A_{\bar m\bar m1}$ and $A_{\bar m\bar m2}$ in the source term Eq.~\eqref{eq:23} of the radial Teukolsky equation Eq.~\eqref{eq:11} are defined as
\begin{equation}
\label{eq:D1}
\begin{split}
A_{nn0} = &\;\frac{P(\pi)}{2\pi \sqrt{2\pi} Q^2} \frac{\Omega^4 C_{nn}}{\sin\theta}
\Bigg\{
2S'(\theta) K^{(0,1)} X_{nn02} 
\\
&\;+ S(\theta) \big( -K^{(0,1)} X_{nn01} + 2K^{(0,1)} X_{nn02}^{(0,1)} 
\\
&\;+ K^{(0,2)} X_{nn02} \big) + K \Big[ -S'(\theta) X_{nn01} 
\\
&\;+ S''(\theta) X_{nn02} + 2S'(\theta) X_{nn02}^{(0,1)} 
\\
&\;+ S(\theta) \big( X_{nn00} - X_{nn01}^{(0,1)} + X_{nn02}^{(0,2)} \big) \Big]
\Bigg\},
\end{split}
\end{equation}
\begin{equation}
\label{eq:D2}
\begin{split}
A_{\bar{m}n0} = &\;\frac{P(\pi)}{2\pi\sqrt{2\pi} Q^3} \frac{\Omega^4 C_{\bar{m}n}}{\sin\theta}
\Bigg\{-2S(\theta) K^{(0,1)} Q^{(1,0)} 
\\
&\;\times\bigl( X1_{\bar{m}n11} + X2_{\bar{m}n11} \bigr)+ Q \Big[
S'(\theta) K^{(1,0)} 
\\
&\;\times\bigl( X1_{\bar{m}n11} + X2_{\bar{m}n11} \bigr)
\\
&\;+ S(\theta) \big(
- K^{(0,1)} X1_{\bar{m}n01}
- K^{(0,1)} X2_{\bar{m}n01}
\\
&\;- K^{(1,0)} X1_{\bar{m}n10}
- K^{(1,0)} X2_{\bar{m}n10}
\\
&\;+ K^{(1,0)} X1_{\bar{m}n11} ^{(0,1)}
+ K^{(1,0)} X2_{\bar{m}n11} ^{(0,1)}
\\
&\;+ K^{(0,1)} X1_{\bar{m}n11} ^{(1,0)}
+ K^{(0,1)} X2_{\bar{m}n11} ^{(1,0)}
\\
&\;+ K^{(1,1)} X1_{\bar{m}n11}
+ K^{(1,1)} X2_{\bar{m}n11}
\big)
\Big]
\\
&\;+ K \Big[
2 Q^{(1,0)} \big(
- S'(\theta) \big( X1_{\bar{m}n11} + X2_{\bar{m}n11} \big)
\\
&\;+ S(\theta) \big( X1_{\bar{m}n10} + X2_{\bar{m}n10}
- X1_{\bar{m}n11} ^{(0,1)}
\\
&\;- X2_{\bar{m}n11} ^{(0,1)} \big)
\big)+ Q \big(
S'(\theta) \big(
- X1_{\bar{m}n01} 
\\
&\;- X2_{\bar{m}n01}
+ X1_{\bar{m}n11} ^{(1,0)}
+ X2_{\bar{m}n11} ^{(1,0)} \big)
\\
&\;+ S(\theta) \big(
X1_{\bar{m}n00} + X2_{\bar{m}n00}
-  X1_{\bar{m}n01} ^{(0,1)}
\\
&\;-  X2_{\bar{m}n01} ^{(0,1)}
-  X1_{\bar{m}n10} ^{(1,0)}
-  X2_{\bar{m}n10} ^{(1,0)}
\\
&\;+  X1_{\bar{m}n11} ^{(1,1)}
+  X2_{\bar{m}n11} ^{(1,1)} \big)
\big)
\Big]
\Bigg\},
\end{split}
\end{equation}
\begin{equation}
\label{eq:D3}
\begin{split}
A_{\bar{m} \bar{m} 0} = &\frac{P(\pi)}{2\pi \sqrt{2\pi} Q^4} 
\frac{\Omega^4 C_{(\bar{m} \bar{m})}}{\sin\theta} S(\theta)
\Bigg\{
Q \Big[ -4K^{(1,0)} \\
&\;\times Q^{(1,0)} X_{\bar{m}\bar{m}20}+ Q \big( -K^{(1,0)} X_{\bar{m}\bar{m}10}
\\
&\;+ 2K^{(1,0)} X_{\bar{m}\bar{m}20}^{(1,0)}
+ K^{(2,0)} X_{\bar{m}\bar{m}20} \big) \Big]
\\
&\;+ K \Big[ 6 (Q^{(1,0)})^2 X_{\bar{m}\bar{m}20}
\\
&\;+ Q^2 \big( X_{\bar{m}\bar{m}00}- X_{\bar{m}\bar{m}10}^{(1,0)}
+ X_{\bar{m}\bar{m}20}^{(2,0)} \big) 
\\
&\;+ 2Q \big( Q^{(1,0)} X_{\bar{m}\bar{m}10}- 2Q^{(1,0)} X_{\bar{m}\bar{m}20}^{(1,0)}
\\
&\;- Q^{(2,0)} X_{\bar{m}\bar{m}20} \big)
\Big]
\Bigg\},
\end{split}
\end{equation}
\begin{equation}
\label{eq:D4}
\begin{split}
A_{\bar{m}n1} = &\;-\frac{P(\pi)}{2\pi\sqrt{2\pi} Q^2} \frac{\Omega^4 C_{(\bar{m}n)}}{\sin\theta} 
\Bigg\{ K^{(0,1)} S(\theta) 
\\
&\;\times \big( X1_{\bar{m}n11} + X2_{\bar{m}n11} \big) 
\\
&\;+ K \Big[ S'(\theta) \big( X1_{\bar{m}n11} + X2_{\bar{m}n11} \big) 
\\
&\;+ S(\theta) \big( -X1_{\bar{m}n10} - X2_{\bar{m}n10} 
\\
&\;+ X1_{\bar{m}n11}^{(0,1)} + X2_{\bar{m}n11}^{(0,1)} \big) \Big] \Bigg\},
\end{split}
\end{equation}
\begin{equation}
\label{eq:D5}
\begin{split}
A_{\bar{m} \bar{m} 1} = &\;-\frac{P(\pi)}{2\pi \sqrt{2\pi} Q^3} 
\frac{\Omega^4 C_{\bar{m} \bar{m}}}{\sin\theta} S(\theta)
\Bigg\{ 2 K^{(1,0)} Q 
\\
&\;\times X_{\bar{m} \bar{m} 20} - K \Big[ 4 Q^{(1,0)} X_{\bar{m} \bar{m} 20} 
\\
&\;+ Q \big( X_{\bar{m} \bar{m} 10} - 2 X_{\bar{m} \bar{m} 20}^{(1,0)} \big) \Big] \Bigg\},
\end{split}
\end{equation}
\begin{equation}
\label{eq:D6}
A_{\bar{m} \bar{m} 2} = 
\frac{P(\pi)}{2\pi \sqrt{2\pi} Q^2} 
\frac{\Omega^4 C_{(\bar{m} \bar{m})}}{\sin\theta} K S(\theta) X_{(\bar{m} \bar{m} 20)},
\end{equation}
with
\begin{equation}
\label{eq:D7}
\begin{split}
K &= \sin\theta \, \Omega^{-3}  8\pi \Sigma \rho_r^{-4}, 
\\
X_{nn00} &= -c_{4b}c_{6b} - c_{4a} c_{6b}^{(0,1)}, \\
X_{nn01} &= -c_{4b}c_{6a} - c_{4a}c_{6b} - c_{4a} c_{6a}^{(0,1)}, 
\\
X_{nn02} &= -c_{4a}c_{6a},
\\
X1_{\bar{m}n00} &= c_{1b}c_{2b} + c_{1a} c_{2b}^{(1,0)}, 
\\
X2_{\bar{m}n00} &= c_{4b}c_{5b} + c_{4a} c_{5b}^{(0,1)}, 
\\
X1_{\bar{m}n01} &= c_{1b}c_{2a} + c_{1a} c_{2a}^{(1,0)}, 
\\
X2_{\bar{m}n01} &= c_{4a}c_{5b}, 
\\
X1_{\bar{m}n10} &= c_{1a}c_{2b}, 
\\
X2_{\bar{m}n10} &= c_{4b}c_{5a} + c_{4a} c_{5a}^{(0,1)},
\\
X1_{\bar{m}n11} &= c_{1a}c_{2a}, 
\\
X2_{\bar{m}n11} &= c_{4a}c_{5a}, 
\\
X_{\bar{m}\bar{m}00} &= -c_{1b}c_{3b} - c_{1a} c_{3b}^{(1,0)},
\\
X_{\bar{m}\bar{m}10} &= -c_{1b}c_{3a} - c_{1a}c_{3b} - c_{1a} c_{3a}^{(1,0)}, 
\\
X_{\bar{m}\bar{m}20} &= -c_{1a}c_{3a},
\end{split}
\end{equation}
where
\begin{equation}
\label{eq:D8}
\begin{split}
c_{1a} &= c_{3a} = c_{5a} = -\frac{Q}{2\Sigma}, 
\\
c_{2a} &= c_{4a} = c_{6a} = \frac{\sqrt{P} \Omega}{\sqrt{2} (r -  i a \cos\theta)}, 
\\
c_{1b} &= 3\gamma - \bar{\gamma} + 4\mu + \bar{\mu} - \frac{ i(a^2\omega + r^2\omega - am)}{2\Sigma}, 
\\
c_{2b} &= 2\alpha - 2\bar{\tau} + \frac{\Omega (m\csc\theta - a\omega\sin\theta)}{\sqrt{2P} (r -  i a \cos\theta)}, 
\\
c_{3b} &= 2\gamma - 2\bar{\gamma} + \bar{\mu} - \frac{ i(a^2\omega + r^2\omega - am)}{2\Sigma}, 
\\
c_{4b} &= 3\alpha + \bar{\beta} + 4\varpi - \bar{\tau} + \frac{\Omega (m\csc\theta - a\omega\sin\theta)}{\sqrt{2P} (r -  i a \cos\theta)}, 
\\
c_{5b} &= 2\gamma + 2\bar{\mu} - \frac{ i(a^2\omega + r^2\omega - am)}{2\Sigma}, 
\\
c_{6b} &= 2\alpha + 2\bar{\beta} - \bar{\tau} + \frac{\Omega (m\csc\theta - a\omega\sin\theta)}{\sqrt{2P} (r -  i a \cos\theta)}.
\end{split}
\end{equation}
The superscript $(N_1,N_2)$ denotes $N_1$ derivatives with respect to $r$ and $N_2$ derivatives with respect to $\theta$. Setting $A=0$ in Eqs.~\eqref{eq:D1}--\eqref{eq:D6} recovers the standard Kerr source coefficients \cite{Sasaki:2003xr}.

\section{Sasaki--Nakamura transformation}
\label{E}
The Sasaki--Nakamura transformation \cite{Sasaki:1981kj,Sasaki:1981sx} maps the homogeneous radial Teukolsky equation~\eqref{eq:24} to the short-range equation~\eqref{eq:25}:
\begin{equation}
\label{eq:E1}
\begin{split}
&Q^2 \partial_r \big( Q^{-1} \partial_r R(r) \big) - V_R R(r) = 0 
\\
&\qquad\qquad\qquad\quad   \Updownarrow
\\
&\partial_{r_*} \big( \partial_{r_*} X(r) \big) - F_1 \partial_{r_*} X(r) - U_1 X(r) = 0
\end{split}\;\;,
\end{equation}
where
\begin{equation}
\label{eq:E2}
\begin{split}
{\partial _{{r_*}}} &= \frac{Q}{{{r^2} + {a^2}}}{\partial _r},\\
F_1 &= \frac{QF}{r^2 + a^2},\\
U_1 &= \frac{QU}{(r^2 + a^2)^2} + G^2 + \partial_{r_*} G - \frac{QGF}{r^2 + a^2},
\end{split}
\end{equation}
and the tortoise coordinate $r_*$ is
\begin{equation}
\label{eq:E3}
\begin{split}
r_* =&\; \frac{1}{2A(1-A^2 r_-^2)(1-A^2 r_+^2)(r_- - r_+)} \\
&\;\times \Bigg\{ (1+a^2A^2)(r_- - r_+)(1+A^2 r_- r_+) \\
&\;\times \ln\left(\frac{1+Ar}{1-Ar}\right) + A \bigg[ -(1+a^2A^2)\\
&\;\times(r_-^2 - r_+^2)\ln(1-A^2 r^2) + 2(a^2+r_-^2) \\
&\;\times(1-A^2 r_+^2)\ln(r - r_-) - 2(a^2+r_+^2) \\
&\;\times(1-A^2 r_-^2)\ln(r - r_+) \bigg] \Bigg\} - 2M \ln(2M).
\end{split}
\end{equation}
Here ${r_ \pm } = M \pm \sqrt {{M^2} - {a^2}}$, and the integration constant is set as $- 2M\ln \left( {2M} \right)$ to make Eq.~\eqref{eq:E3} return the Kerr result when $A=0$. The functions $F$, $U$ and $G$ in Eq.~\eqref{eq:E2} are defined as
\begin{equation}
\label{eq:E4}
\begin{split}
F =&\; \frac{\partial_r \eta}{\eta}, 
\\
U =&\; V_R + \frac{Q^2}{\beta} \bigg[ \partial_r \Big( 2\alpha + \frac{\partial_r \beta}{Q} \Big) 
\\
&\;- \frac{\partial_r \eta}{\eta} \Big( \alpha + \frac{\partial_r \beta}{Q} \Big) \bigg],
\\
G =&\; -\frac{\partial_r Q}{r^2 + a^2} + \frac{rQ}{(r^2 + a^2)^2},
\end{split}
\end{equation}
where
\begin{equation}
\label{eq:E5}
\begin{split}
\alpha =&\; f \bigg\{ \frac{Q (r^2 + a^2)}{g h} J_- \Big[ \frac{h}{J_-} \big( \frac{g}{r^2 + a^2} \big) \Big] + V_R \bigg\},
\\
\beta =&\; f Q^2 (r^2 + a^2) \bigg[ \frac{1}{Q g} J_- \Big( \frac{Q g}{r^2 + a^2} \Big) 
\\
&\;+ \frac{1}{g h} J_- \Big( \frac{g h}{r^2 + a^2} \Big) \bigg],
\\
\eta =&\; \alpha \Big( \alpha + \frac{\partial_r \beta}{Q} \Big) - \frac{\beta}{Q} \Big( \partial_r \alpha + \frac{\beta}{Q^2} V_R \Big),
\end{split}
\end{equation}
with
\begin{equation}
\label{eq:E6}
\begin{split}
f &= 1, \quad h = 1, \quad g = \frac{r^2 + a^2}{r^2},
\\
J_{\pm} &= \partial_r \pm i \frac{(r^2 + a^2)\omega - am}{Q}. 
\end{split}
\end{equation}
The forward transformation from $R(r)$ to $ X(r)$ is
\begin{equation}
\label{eq:E7}
X(r) = \sqrt{Q^{-2} \big( r^2 + a^2 \big)} 
\Big( \alpha R(r) + \beta Q^{-1} \partial_r R(r) \Big),
\end{equation}
and the inverse transformation from $X(r)$ to $ R(r)$ is
\begin{equation}
\label{eq:E8}
\begin{split}
R(r) =&\; \frac{1}{\eta} \Bigg[
\big( \alpha + Q^{-1} \partial_r \beta \big) 
\frac{X(r)}{\sqrt{Q^{-2} (r^2 + a^2)}} \\
&\;- \beta Q^{-1} \partial_r \bigg( \frac{X(r)}{\sqrt{Q^{-2} (r^2 + a^2)}} \bigg)
\Bigg].
\end{split}
\end{equation}


\twocolumngrid
\bibliography{REF} 
\bibliographystyle{apsrev4-2}

\end{document}